\documentclass[nofootinbib,preprint,tightenlines]{revtex4}

\usepackage[utf8]{inputenc}
\usepackage{amssymb,amsthm,amsmath,amstext,amsbsy}
\usepackage{url}
\usepackage{nicefrac}
\usepackage{graphicx}
\usepackage{hyperref}

\newcommand{\ie}{\textit{i.e.}}
\newcommand{\eg}{\textit{e.g.}}
\newcommand{\cf}{\textit{cf.}}

\newcommand{\gsim}{\gtrsim}
\newcommand{\lsim}{\lesssim}

\newcommand{\LO}{\text{LO}}
\newcommand{\NLO}{\text{NLO}}
\newcommand{\NNLO}{\text{N$^2$LO}}

\newcommand{\MeV}{\ensuremath{\mathrm{MeV}}}
\newcommand{\fm}{\ensuremath{\mathrm{fm}}}

\newcommand{\dd}{\mathrm{d}}
\newcommand{\ii}{\mathrm{i}}
\newcommand{\vD}{\boldsymbol{D}}
\newcommand{\hc}{\mathrm{h.c.}}
\newcommand{\OO}{\mathcal{O}}

\newcommand{\eps}{\varepsilon}
\newcommand{\one}{\mathbf{1}}
\newcommand{\diag}{\mathrm{diag}}
\newcommand{\leviciv}{\epsilon}
\newcommand{\vdelta}{\delta^{(3)}}
\newcommand{\fvdelta}{\delta^{(4)}}

\newcommand{\bra}[1]{\left\langle #1\right|}
\newcommand{\ket}[1]{\left|#1\right\rangle}

\newcommand{\mathtext}[1]{\ \ \text{#1}\ \ }

\newcommand{\EB}{E_\mathrm{B}}

\newcommand{\MN}{M_N}

\newcommand{\Mpi}{M_\pi}

\newcommand{\yd}{y_d}
\newcommand{\yt}{y_t}

\newcommand{\sigmad}{\sigma_d}
\newcommand{\sigmat}{\sigma_t}

\newcommand{\gamd}{\gamma_d}
\newcommand{\rd}{\rho_d}
\newcommand{\rnt}{\rho_t}
\newcommand{\rnC}{r_C}

\newcommand{\sss}{\mathrm{s}}
\newcommand{\ccc}{\mathrm{c}}
\newcommand{\fff}{\mathrm{full}}
\newcommand{\ddd}{\mathrm{diff}}

\newcommand{\Tgen}{\mathcal{T}}
\newcommand{\TS}{\Tgen_\sss}
\newcommand{\TC}{\Tgen_\ccc}
\newcommand{\TF}{\Tgen_\fff}

\newcommand{\TSq}{\TS^\mathrm{q}}
\newcommand{\TCq}{\TC^\mathrm{q}}
\newcommand{\TFq}{\TF^\mathrm{q}}

\newcommand{\TSda}{\TS^\mathrm{d,a}}
\newcommand{\TSdb}{\TS^\mathrm{d,b}}
\newcommand{\TFda}{\TF^\mathrm{d,a}}
\newcommand{\TFdb}{\TF^\mathrm{d,b1}}
\newcommand{\TFdc}{\TF^\mathrm{d,b2}}

\newcommand{\Bgen}{\mathcal{B}}
\newcommand{\BS}{\Bgen_\sss}
\newcommand{\BSda}{\BS^\mathrm{d,a}}
\newcommand{\BSdb}{\BS^\mathrm{d,b1}}
\newcommand{\BSdc}{\BS^\mathrm{d,b2}}

\newcommand{\KS}{K_\sss}
\newcommand{\KC}{K_\ccc}

\newcommand{\KCd}{\KC^{(d)}}
\newcommand{\KCt}{\KC^{(t)}}
\newcommand{\KCdt}{\KC^{(d,t)}}

\newcommand{\deltaC}{\delta_\ccc}
\newcommand{\deltaF}{\delta_\fff}

\newcommand{\dq}[1]{\!\!\frac{\mathrm{d}^3#1}{(2\pi)^3}}
\newcommand{\dn}[1]{\!\frac{\mathrm{d}#1_0}{2\pi}}
\newcommand{\dfq}[1]{\!\!\frac{\mathrm{d}^4#1}{(2\pi)^4}}

\newcommand{\ddq}{\dq{q}}

\newcommand{\vk}{\mathbf{k}}
\newcommand{\vp}{\mathbf{p}}
\newcommand{\vq}{\mathbf{q}}
\newcommand{\vP}{\mathbf{P}}
\newcommand{\vZero}{\mathbf{0}}

\newcommand{\sktilde}[1]{\widetilde{#1}}

\newcommand{\skrowspace}{0.5em}

\begin{document}

\title{Low-energy $p$--$d$ scattering and $^3$He in pionless EFT}

\author{Sebastian König}
\email{koenig@hiskp.uni-bonn.de}
\affiliation{Helmholtz-Institut für Strahlen- und Kernphysik (Theorie)\\
and Bethe Center for Theoretical Physics, Universität Bonn, 53115 Bonn,
Germany\\}

\author{H.-W. Hammer}
\email{hammer@hiskp.uni-bonn.de}
\affiliation{Helmholtz-Institut für Strahlen- und Kernphysik (Theorie)\\
and Bethe Center for Theoretical Physics, Universität Bonn, 53115 Bonn,
Germany\\}

\date{\today}

\begin{abstract}
We calculate low-energy proton--deuteron scattering in the framework of pionless
effective field theory. In the quartet channel, we calculate the elastic
scattering phase shift up to next-to-next-to-leading order in the power
counting. In the doublet channel, we perform a next-to-leading order
calculation. We obtain good agreement with the available phase shift analyses
down to the scattering threshold. The phase shifts in the region of
non-perturbative Coulomb interactions are calculated by using an optimised
integration mesh. Moreover, the Coulomb contribution to the $^3$He--$^3$H
binding energy difference is evaluated in first order perturbation theory. We
comment on the implications of our results for the power counting of subleading
three-body forces.
\end{abstract}

\maketitle

\section{Introduction}

Although Quantum Chromodynamics (QCD) is widely accepted as the underlying
theory of strong interactions, \textit{ab initio} calculations of nuclear
properties in Lattice QCD remain a large theoretical
challenge~\cite{Beane:2010em}.  In nuclear physics, the relevant degrees of
freedom are pions and nucleons, and much of the computational effort in such a
calculation would be required for generating the correct degrees of freedom from
quarks and gluons rather than their interactions. Traditionally,
nucleon--nucleon interactions are described via phenomenological nuclear forces
fitted to scattering data.  Effective field theory (EFT) provides a powerful
method to construct nuclear forces with a direct connection to QCD in a
systematic, model-independent
way~\cite{Beane:2000fx,Bedaque:2002mn,Epelbaum:2008ga}.

For very low energies and momenta $p\lsim\Mpi$, the non-analyticities from pion
exchange cannot be resolved and one can hence use an EFT including only
short-range contact interactions between
nucleons~\cite{Kaplan:1998we,vanKolck:1998bw}. This theory is constructed to
reproduce the effective range expansion~\cite{Bethe:1949yr} in the two-body
system and recovers Efimov's universal approach to the three-nucleon
problem~\cite{Efimov:1981aa,Hammer:2010kp}. An advantage of the EFT formulation
is that it can be extended to higher-body systems and external currents in a
straightforward way.

The extension of this EFT to include the long-range Coulomb interaction was
first discussed by Kong and Ravndal for the proton--proton
channel~\cite{Kong:1998sx,Kong:1999sf}. In Ref.~\cite{Ando:2007fh}, this
analysis was extended to next-to-next-to-leading order. A renormalisation group
analysis of proton--proton  scattering in a distorted wave basis was carried out
in Refs.~\cite{Barford:2002je,Ando:2008jb}. Moreover, this theory was applied
to proton--proton fusion in Refs.~\cite{Kong:2000px,Ando:2008va}.
An extension of this formalism to three charged particles would be important
for the possible interpretation of the Hoyle state in $^{12}$C as an Efimov
state of $\alpha$ particles~\cite{Efimov:1970ab} and the cluster EFT for halo 
nuclei~\cite{Bertulani:2002sz}.

In this work, we are interested in the simpler problem of a three-body system
with two charged particles. Close to threshold, the  Coulomb interaction is
strong. Its long-range nature requires special care in the non-perturbative
treatment using momentum space integral equations. At higher energies, the
Coulomb interaction becomes perturbative. Rupak and Kong have formulated a power
counting for the Coulomb contributions in the quartet channel of
proton--deuteron ($p$--$d$) scattering. They calculated the phase shifts to
next-to-next-to-leading order (\NNLO) in the pionless EFT and included Coulomb
effects to next-to-leading order (\NLO)~\cite{Rupak:2001ci}. However, they were
not able to extend their calculation to the threshold region below
center-of-mass momenta of 20~\MeV. They did not consider the doublet channel and
the $^3$He bound state. A leading order calculation of the $^3$He nucleus
including non-perturbative Coulomb interactions was recently presented by Ando
and Birse~\cite{Ando:2010wq}. Including isospin breaking effects in the
nucleon--nucleon scattering lengths, they obtain a surprisingly accurate
description of the $^3$He--$^3$H binding energy difference but they did not
consider scattering observables. A similar study to \NLO\ in the pionless EFT
was carried out using the resonating group method~\cite{Kirscher:2009aj}. Their
results do not include isospin breaking and are consistent with other
determinations of the $^3$He--$^3$H binding energy difference.

In this paper, we focus on $p$--$d$ scattering observables in the quartet
and doublet channels. We extend the power counting by Rupak and Kong for the
Coulomb contribution to the doublet channel. By using a special integration
mesh, we are able to calculate the phase shifts in both channels down to momenta
of order 3~\MeV. We also provide a perturbative evaluation of the Coulomb
contribution to the $^3$He--$^3$H binding energy difference.

\section{Formalism}

In this section, we briefly summarise the formalism required for calculating
$p$--$d$ scattering in the pionless theory. More technical details can, \eg, be
found in Refs.~\cite{Rupak:2001ci,Bedaque:1999ve,Bedaque:2002yg,Braaten:2004rn,
Gabbiani:1999yv}.

\subsection{Effective Lagrangian}

We use the effective Lagrangian
\begin{multline}
 \mathcal{L} = N^\dagger\left(\ii D_0+\frac{\vD^2}{2\MN}\right)N
 -d^{i\dagger}\left[\sigmad+\left(\ii D_0+\frac{\vD^2}{4\MN}\right)\right]d^i
 -t^{A\dagger}\left[\sigmat+\left(\ii D_0+\frac{\vD^2}{4\MN}\right)\right]t^A
 \\[0.2cm]
 +\yd\left[d^{i\dagger}\left(N^T P^i_d N\right)+\hc\right]
 +\yt\left[t^{A\dagger}\left(N^T P^A_t N\right)+\hc\right]
 +\mathcal{L}_\mathrm{photon}+\mathcal{L}_3 \,,
\label{eq:L-Nd}
\end{multline}
with the nucleon field $N$ and two dibaryon fields $d^i$ (with spin 1 and
isospin 0) and $t^A$ (with spin 0 and isospin 1), corresponding to the deuteron
and the spin-singlet virtual bound state in S-wave nucleon--nucleon scattering.
Both dibaryon fields are formally ghosts since their kinetic terms have a 
negative sign. This is required to avoid the Wigner bound and reproduce the
positive value of the effective range with short-range
interactions~\cite{Phillips:1996ae}. Spin and isospin degrees of freedom are
included by treating the field $N$ as a doublet in both spaces, but for
notational convenience we have suppressed the spin and isospin  indices of $N$.
The projection operators,
\begin{equation}
 P^i_d = \frac{1}{\sqrt8}\,\sigma^2\sigma^i\tau^2 \mathtext{,}
 P^A_t = \frac{1}{\sqrt8}\,\sigma^2\tau^2\tau^A \,,
\label{eq:L-Nd-Projectors}
\end{equation}
with $\vec{\sigma}$ ($\vec{\tau}$) operating in spin (isospin) space, project
out the $^3S_1$ and $^1S_0$ nucleon--nucleon partial waves, respectively.

The covariant derivative
\begin{equation}
 D_\mu = \partial_\mu + \ii eA_\mu\cdot\hat{Q} \,,
\label{eq:D-mu}
\end{equation}
where $\hat{Q}$ is the charge operator, includes the coupling to the 
electromagnetic field. Furthermore, we have the kinetic and gauge fixing terms
for the photons,
\begin{equation}
 \mathcal{L}_\mathrm{photon} = -\frac14 F_{\mu\nu}F^{\mu\nu} -\frac{1}{2\xi}
 \left(\partial_\mu A^\mu-\eta_\mu\eta_\nu\partial^\nu A^\mu\right)^2 \,,
\label{eq:L-photon}
\end{equation}
of which we only keep contributions from Coulomb photons. These correspond to a
static Coulomb potential between charged particles, but for convenience we
introduce Feynman rules for a Coulomb photon propagator,
\begin{equation}
 \ii\Delta_{\mathrm{Coulomb}}(k) = \frac{\ii}{\vk^2+\lambda^2} \,,
\label{eq:FR-Coulomb-Propagator}
\end{equation}
which we draw as a wavy line, and factors $(\pm\ii e\cdot\hat{Q})$ for the
vertices.\footnote{Due to the sign convention chosen in the
Lagrangian~\eqref{eq:L-Nd}, dibaryon--photon vertices get an additional minus
sign.} Following~\cite{Rupak:2001ci}, we have introduced a small photon mass
$\lambda$ to regulate the singularity of the propagator at zero momentum
transfer. As we will discuss later on, this regulator will be removed by
numerical extrapolation back to vanishing photon mass.

In the doublet-channel of the three-nucleon system, a three-body contact
interaction is required for renormalisation at leading
order~\cite{Bedaque:1999ve}. It can be written as
\begin{multline}
 \mathcal{L}_3=-\MN\frac{ H(\Lambda)}{\Lambda^2}\Bigg(\yd^2N^\dagger(\vec{d}
 \cdot\vec{\sigma})^\dagger(\vec{d}\cdot\vec{\sigma})N+\yt^2N^\dagger(\vec{t}
 \cdot\vec{\tau})^\dagger(\vec{t}\cdot\vec{\tau})N\\
 +\frac13\yd\yt\left[N^\dagger(\vec{d}\cdot\vec{\sigma})^\dagger(\vec{t}
 \cdot\vec{\tau} )N+\hc\right]
 \Bigg) \,,
\label{eq:L-3}
\end{multline}
where $\Lambda$ is a momentum cutoff applied in the three-body equations
discussed below and $H(\Lambda)$ a known log-periodic function of the cutoff
that depends on a three-body parameter $\Lambda_*$.

\subsection{Dibaryon propagators}

\begin{figure}[htbp]
\centering
\includegraphics[clip]{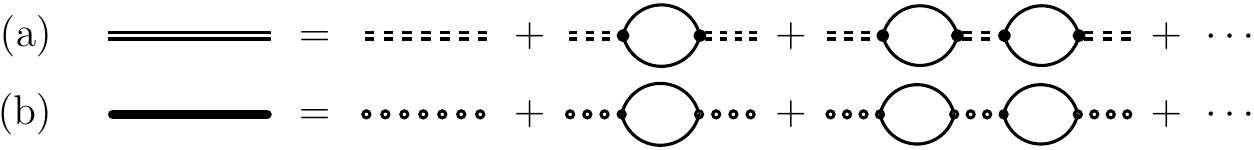}
\caption{Full dibaryon propagators in (a) the $^3S_1$ state (\ie~the deuteron)
and (b) the $^1S_0$ state.}
\label{fig:DibaryonProp}
\end{figure}

In the strong sector, we adopt the standard power counting for large 
S-wave scattering length~\cite{Kaplan:1998we,vanKolck:1998bw}.
A nucleon bubble together with a bare dibaryon propagator scales as $\OO(1)$.
The bare dibaryon propagators therefore are dressed by nucleon bubbles to all
orders. The resulting geometric series for the full propagators are shown
in Fig.~\ref{fig:DibaryonProp}.

For convenience, we also resum the effective range corrections. If desired, the
perturbative expressions can always be obtained by re-expanding the propagators.
We do not go into the details of the calculations here and simply quote the
results for the renormalised propagators, which we obtain by demanding that the
effective range expansions
\begin{equation}
 k\cot\delta_d = -\gamd + \frac{\rd}{2}(k^2+\gamd^2)+\,\cdots
\label{eq:ER-d}
\end{equation}
around the deuteron pole, and
\begin{equation}
 k\cot\delta_t = -\frac1{a_t} + \frac{\rnt}{2}k^2+\,\cdots
\label{eq:ER-t}
\end{equation}
for the singlet channel are reproduced. In writing Eq.~(\ref{eq:ER-t}) we have
used that $\rnt=r_{0t}$ to the order we are working. This yields the expressions
\begin{equation}
 \ii\Delta^{ij}_d(p) = -\frac{4\pi\ii}{\MN\yd^2}
 \cdot\frac{\delta^{ij}}{-\gamd+\sqrt{\frac{\vp^2}{4}-\MN p_0-\ii\eps}
 -\frac{\rd}{2}\left(\frac{\vp^2}{4 }-\MN p_0-\gamd^2\right)} \,,
\label{eq:Prop-d-High}
\end{equation}
and analogously
\begin{equation}
 \ii\Delta^{AB}_t(p) = -\frac{4\pi\ii}{\MN\yt^2}
 \cdot\frac{\delta^{AB}}{-\frac1{a_t}+\sqrt{\frac{\vp^2}{4}-\MN p_0-\ii\eps}
 -\frac{\rnt}{2}\left(\frac{\vp^2}{ 4}-\MN p_0\right)}
\label{eq:Prop-t-High}
\end{equation}
for the spin-singlet dibaryon. These expressions are valid to \NNLO. At leading
order, effective range corrections are not included and the dibaryon kinetic
terms do not contribute. The corresponding propagators are obtained by setting
$\rnt=0$ and $\rd=0$ in Eqs.~(\ref{eq:Prop-d-High}) and (\ref{eq:Prop-t-High}).

The deuteron wave function renormalisation constant is given as the residue at
the bound state pole:
\begin{equation}
 Z_0^{-1} = \ii\frac{\partial}{\partial p_0}
 \left.\frac{1}{\ii\Delta_d(p)}\right|_{p_0 =-\frac{\gamd^2}{\MN},\,\vp=0} \,.
\label{eq:Z0-LO}
\end{equation}

\subsection{Coulomb contributions in the proton--proton system}

\begin{figure}[htbp]
\centering
\includegraphics[clip]{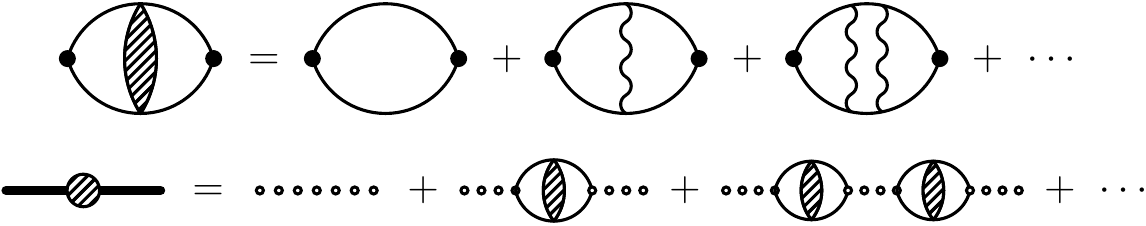}
\caption{Dressed nucleon bubble and full singlet dibaryon propagator in the 
$pp$-channel.}
\label{fig:DressedBubble}
\end{figure}

The Coulomb interaction breaks the isospin symmetry that is implicit in the
dibaryon propagators from the previous subsection. For the $pp$-part of the
singlet dibaryon we can also have Coulomb photon exchanges inside the nucleon
bubble. These can be resummed to all orders, yielding a dressed nucleon
bubble~\cite{Kong:1998sx,Kong:1999sf}, which is then used to calculate the full
singlet dibaryon propagator in the $pp$-channel, as shown in 
Fig.~\ref{fig:DressedBubble}.

The result for the leading order propagator is~\cite{Ando:2010wq}
\begin{equation}
 \ii\Delta^{AB}_{t,pp}(p) = -\frac{4\pi\ii}{\MN\yt^2}
 \cdot\frac{\delta^{AB}}{-1/a_C-2\kappa H(\kappa/p')} \mathtext{,}
 \kappa=\frac{\alpha\MN}{2}
\label{eq:Prop-t-pp-Low}
\end{equation}
with
\begin{equation}
 p' = \ii\sqrt{{\vp^2}/{4}-\MN p_0-\ii\eps}
\end{equation}
and
\begin{equation}
 H(\eta) = \psi(\ii\eta)+\frac{1}{2\ii\eta}-\log(\ii\eta) \,,
\end{equation}
where $\psi$ denotes the logarithmic derivative of the $\Gamma$-function.
Effective range corrections can be included in the same way as described above.

\subsection{Power counting}
\label{sec:Power-Counting}

The power counting of pionless effective field theory has been extensively
discussed in the literature (see the 
reviews~\cite{Beane:2000fx,Bedaque:2002mn,Epelbaum:2008ga} and references
therein). We will thus be rather brief on this subject here. We will, however,
elaborate a bit on the power counting for the Coulomb sector of the theory, as
it was introduced in~\cite{Rupak:2001ci}.

\subsubsection{Strong sector}

The low-energy scale $Q$ of the theory is set by the deuteron binding momentum
$\gamd\sim45~\MeV$. We can formally count the external momenta $k,p$ to be of
the same order. Since we are working in a theory without explicit pions, the
natural ultraviolet cutoff of our theory is of the order of the pion mass,
$\Lambda\sim \Mpi$. Which cutoff is best to use in practice depends on whether
one discusses the quartet-channel system (where short-range effects are
suppressed by the Pauli principle), or the doublet-channel system. In the first
case one finds that already $\Lambda\approx140~\MeV$ is sufficient for an
accurate description, whereas in the latter case we have to set the cutoff to a
few hundred \MeV\ to reach convergence. The combination of the two scales yields
the expansion parameter $\OO(Q/\Lambda)$.

A further relevant scale in our system is the nucleon mass $\MN$. It appears
explicitly in kinetic energies, which scale as $\OO(Q^2/\MN)$. As a consequence,
the nucleon propagator scales as $\OO(\MN/Q^2)$ and the loop integration measure
$\dd^3q\,\dd q_0$ scales as $\OO(Q^5/\MN)$. We assume
$\yd^2\sim\yt^2\sim\Lambda/\MN^2$ for the nucleon--dibaryon coupling constants
and $\sigmad\sim\sigmat\sim Q\Lambda/\MN$ for the bare dibaryon propagator
constants.

\subsubsection{Including Coulomb photons}

From the form of the Coulomb potential in momentum space,
\begin{equation}
 V_\ccc(q) \sim \frac{\alpha}{q^2} \,,
\end{equation}
it is clear that Coulomb contributions are dominant for small momentum
transfers. As noted in~\cite{Rupak:2001ci}, they enter $\sim\alpha\MN/q$. This
behaviour is not captured by the power counting for the strong sector. Hence,
when one wants to perform calculations including Coulomb effects for small
external momenta, one can no longer assume that all momenta scale with
$Q\sim\gamd$. Instead, one has to keep track of the new scale introduced by the
external momenta separately. We generically denote this scale by $p$ and assume
$p\ll Q$ for the power counting. As noted in~\cite{Rupak:2001ci}, this means
that we make a simultaneous expansion in \emph{two} small parameters $Q/\Lambda$
and $p/(\alpha\MN)$. For $p\gsim Q$, the Coulomb contributions are small and the
results in both schemes agree.

With this modified counting, it is not straightforward to deduce the scaling of
loops anymore. Kinetic energies always scale like $Q$, so the scaling of $\dd
q_0$ and the nucleon propagator is not modified in the presence of Coulomb
effects. However, where we could simply assume that all loop momenta scale like
$Q$ before, we now have to check first which contribution is picked up (or
rather enhanced) after carrying out the $\dd q_0$-integral. In general, we have
that
\begin{enumerate}
 \item the loop integration measure $\dd^3q$ scales as $q^3$, and
 \item the photon propagator scales as $1/q^2$,
\end{enumerate}
where either $q\sim Q$ or $q\sim p$. These rules will become more transparent
when we apply them to deduce the scaling of the diagrams shown below.

\subsubsection{Selected diagrams}
\label{sec:Selected-Diagrams}

\begin{figure}[htbp]
\centering
\includegraphics[clip]{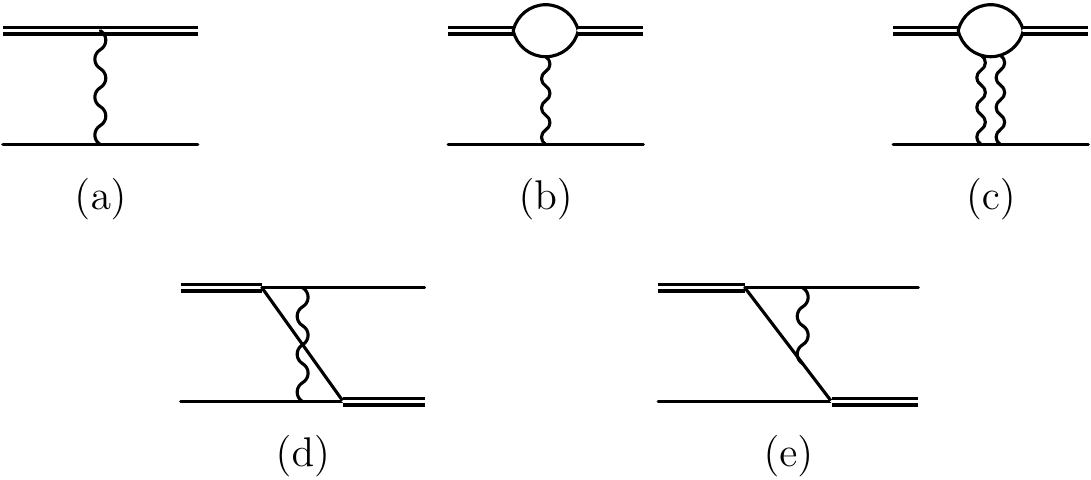}
\caption{Diagrams for $p$--$d$ scattering involving the exchange of Coulomb
photons.}
\label{fig:CoulombDiagrams}
\end{figure}

In this subsection, we discuss several diagrams contributing to $p$--$d$
scattering that include Coulomb photons (see Fig.~\ref{fig:CoulombDiagrams}).
For the discussion we always assume $p \ll Q$. Diagram (a) simply scales as
$\alpha/p^2$. The diagram (b) is enhanced relative to (a) by a factor
$\Lambda/Q$ from the nucleon bubble and hence gives the leading-order Coulomb
contribution. Diagram (a) enters only at \NLO\ since the dibaryon kinetic energy
operators, which generate the direct coupling of the photons to the dibaryons,
enter only as effective range corrections.

Diagram (c) contains two loops, which have to be analysed separately. The upper
nucleon bubble does not contain any photon propagators, so there all momenta
scale as $Q$. In the lower loop, momenta certainly scale $\sim p$ due to the two
photon propagators that involve the external momentum. Including the remaining
nucleon propagator, which cancels the contribution from the integration measure,
we are left with a total scaling $\sim\alpha^2\MN\Lambda/(Q^3p)$ for diagram
(c). This means that compared to diagram (b) it is suppressed by a factor
$\alpha\MN p/Q^2$.
For diagrams of the form (c) with more than two photons attached to the bubble
we simply quote the results from~\cite{Rupak:2001ci}. The diagram with three
photons could contribute with a factor $\sim\log(p/Q)$, whereas the diagrams
with $n>3$ photons attached to the bubble are even infrared finite and
suppressed by factors $\alpha^n$. Following~\cite{Rupak:2001ci}, we neglect them
all and also the logarithmically-scaling diagram with three photons (which is
already small for $p\ge1~\MeV$).

Diagram (d) is a little ambiguous since \textit{a priori} it is not clear
whether the loop momentum should scale $\sim Q$ or $\sim p$.
In~\cite{Rupak:2001ci}, the first alternative is chosen, yielding that compared
to the same diagram without the photon it is suppressed by a factor
$\alpha\MN/Q$.  A direct numerical calculation shows that it is a seven-percent
effect at threshold.

The diagram (e) obviously is irrelevant for the quartet-channel system (there
are never two protons in the dibaryon), but, at least in principle, it can play
a role in the doublet-channel system. The power counting, however, yields
the same suppression factor as for diagram (c), only in this case the scaling of
the loop momentum is not ambiguous. A direct numerical evaluation yields that at
threshold it is a 15\% contribution (again compared to the simple
nucleon-exchange diagram without the photon). We take this value as the
\textit{a priori} theoretical uncertainty of our doublet-channel calculation.

\medskip
The bottom line of the discussion above is that, as done in~\cite{Rupak:2001ci},
we iterate the diagrams (a) and (b) to all orders and do not include any of the
other diagrams shown in Fig.~\ref{fig:CoulombDiagrams}. The claim is that this
procedure is adequate for both the quartet-channel and the doublet-channel
system. The Coulomb effects are thus included at \NLO\ accuracy in our 
calculation.

\section{Scattering equations}

\subsection{Quartet channel}

We start with a review of $N$--$d$ quartet-channel scattering, where the spin 1
of the deuteron and the spin $\nicefrac12$ of the nucleon couple to a total spin
of $\nicefrac32$. Since this coupling is only possible when the spins of all
three nucleons taking part in the reaction are aligned, the Pauli principle
applies. Hence, the system is rather insensitive to short-range physics.
Furthermore, only the dibaryon field representing the deuteron can appear in the
intermediate state.

\subsubsection*{Neutron--deuteron system}

\begin{figure}[htbp]
\centering
\includegraphics[clip]{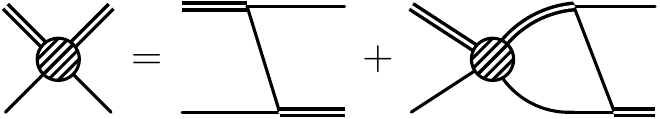}
\caption{Integral equation for the strong scattering amplitude $\TS$ in the
quartet channel.}
\label{fig:nd-IntEq-Q}
\end{figure}

In Fig.~\ref{fig:nd-IntEq-Q} we show a diagrammatic representation of the strong
(neutron--deuteron) scattering amplitude $\TS$, which does not include any
Coulomb effects. It is projected onto the spin quartet channel by setting
$i=(1-\ii 2)/\sqrt{2}$ and $j=(1+\ii 2)/\sqrt{2}$ for the in- and outgoing
deuteron spin indices, respectively, and $a=b=2$ to select the neutron. After
furthermore projecting onto S-waves, we get
\begin{multline}
 \TSq(E;k,p)=-\frac{\MN\yd^2}{kp}Q\left(\frac{k^2+p^2-\MN E-\ii\eps}{kp}\right)
 \\+\frac{1}{2\pi^2}\int_0^\Lambda\ddq\,q^2\,\TSq(E;k,q)\,
 \Delta_d\left(E-\frac{q^2}{2\MN},q\right)\\
 \times\frac{\MN\yd^2}{qp}Q\left(\frac{q^2+p^2-\MN E-\ii\eps}{qp}\right) \,,
\label{eq:nd-IntEq-Q}
\end{multline}
where $k$ and $p$ are the incoming and outgoing momenta of the particles 
in the center-of-mass
frame, and
\begin{equation}
 Q(a) = \frac{1}{2}\int_{-1}^1\frac{\dd x}{x+a}
 = \frac{1}{2}\log\left(\frac{a+1}{a-1}\right) \,.
\label{eq:Q}
\end{equation}
More details on the derivation of this equation and the required projections
can be found in Appendix~\ref{sec:ScattEq-Details}. We have introduced a cutoff
$\Lambda$ to regularise the loop integral, which is particularly convenient for
a numerical treatment of the equation. Strictly, however, this regulator is only
required for the full amplitude including Coulomb interactions and is applied
here for convenience only. After setting the energy $E$ to the total
center-of-mass energy,
\begin{equation}
 E_k = \frac{3k^2}{4\MN}-\frac{\gamd^2}{\MN} \,,
\label{eq:E-on-shell}
\end{equation}
the equation is solved numerically with standard linear algebra routines after
discretising the integrals. From the result we then obtain the S-wave scattering
phase shift
\begin{equation}
 \delta(k) = \frac{1}{2\ii}
 \log\left(1+\frac{2\ii k\MN}{3\pi} Z_0\Tgen(E_k;k,k)\right) \,,
\label{eq:delta}
\end{equation}
which can be compared to experimental data.\par
\medskip
In order to simplify the expressions in the following sections, we introduce a
short-hand notation for the scattering equations. Defining
\begin{equation}
 D_{d,t}(E;q)\equiv\Delta_{d,t}\left(E-\frac{q^2}{2\MN},q\right)
\label{eq:D-d-t-LO}
\end{equation}
and
\begin{equation}
 \KS(E;k,p) \equiv \frac{1}{kp}\;
 Q\left(\frac{k^2+p^2-\MN E-\ii\eps}{kp}\right) \,,
\label{eq:KS}
\end{equation}
along with the operation
\begin{equation}
 A \otimes B \equiv \frac1{2\pi^2}
 \int_0^\Lambda\dd q\,q^2\,A(\ldots,q)B(q,\ldots) \,,
\end{equation}
we find that we can write \eqref{eq:nd-IntEq-Q} as
\begin{equation}
 \TSq = -\MN\yd^2\,\KS + \TSq \otimes \left[\MN\yd^2\,D_d \KS\right] \,,
\end{equation}
where we have omitted all arguments.

\subsubsection*{Proton--deuteron system}

\begin{figure}[htbp]
\centering
\includegraphics[clip]{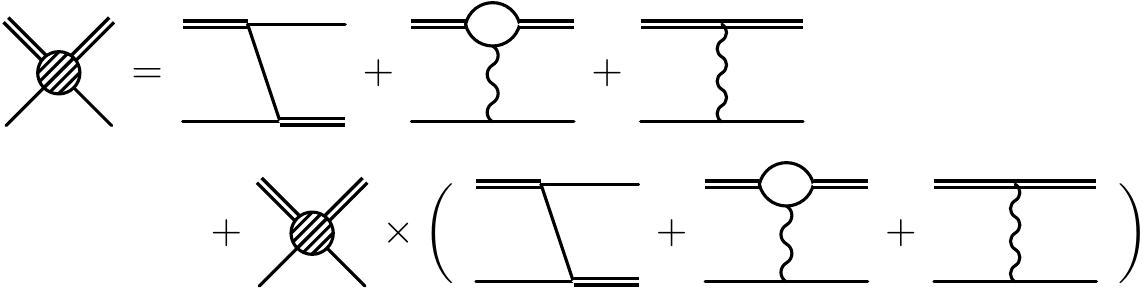}
\caption{Integral equation for the full (\ie~strong + Coulomb) scattering
amplitude $\TF$ in the quartet channel.}
\label{fig:pd-IntEq-Q}
\end{figure}
%
\begin{figure}[htbp]
\centering
\includegraphics[clip]{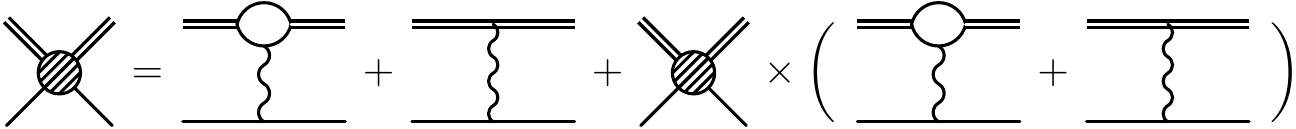}
\caption{Integral equation for the Coulomb scattering amplitude $\TC$.}
\label{fig:pd-IntEq-Coulomb}
\end{figure}

In order to include Coulomb effects and hence discuss proton--deuteron
scattering we follow~\cite{Rupak:2001ci} and define a full scattering amplitude
$\TF$ (see Fig.~\ref{fig:pd-IntEq-Q}) that includes both strong and Coulomb
interactions and a pure Coulomb scattering amplitude $\TC$ (see
Fig.~\ref{fig:pd-IntEq-Coulomb}). After spin-, isospin- and S-wave projection we
find the integral equations
\begin{equation}
 \TFq = -\MN\yd^2\left(K_s -\frac12\KCd\right)
  + \TFq \otimes \left[\MN\yd^2\,D_d\left(K_s -\frac12\KCd\right)\right]
\label{eq:pd-IntEq-Q-full}
\end{equation}
and
\begin{equation}
 \TCq = \frac{\MN\yd^2}2\KCd
 - \TCq \otimes \left[\frac{\MN\yd^2}2 D_d \KCd\right]
\label{eq:pd-IntEq-Q-c}
\end{equation}
with
\begin{equation}
 \KCdt(E;k,p) = \frac{\alpha\MN}{2kp}\;
 Q\left(-\frac{k^2+p^2+\lambda^2}{2kp}\right)
 \left(\frac{1}{|\gamd|}-\rho_{d,t}\right) \,.
\label{eq:KCd}
\end{equation}
After solving the individual equations, we calculate the phase shifts $\deltaF$
and $\deltaC$ according to~\eqref{eq:delta}. The final result, which we will
compare to experimental data, is the Coulomb-subtracted phase shift
\cite{Jackson:1950zz,Harrington:1965,Rupak:2001ci},
\begin{equation}
 \delta_\ddd(k)\equiv\deltaF(k)-\deltaC(k) \,.
\label{eq:delta-diff}
\end{equation}
Note that in the integral equations above an artificial dependence on the bare
coupling constant $y_d$ is kept for notational convenience. In all observables,
this dependence drops out.

\subsection{Doublet channel}

We now go on to the doublet channel, where the spins of the nucleon and the
deuteron couple to a total spin of $\nicefrac12$. The spin-singlet dibaryon
can now appear in the intermediate state, which leads to two coupled amplitudes
that differ in the type of the outgoing dibaryon. In contrast to the quartet
channel, the three nucleon spins no longer need to be aligned in the same
direction, which means that a non-derivative three-nucleon interaction is no 
longer prohibited by the Pauli principle. This channel is expected to be more
sensitive to short-range physics in general, and in fact the three-body
interaction~(\ref{eq:L-3}) is \emph{needed} at leading order to ensure correct
renormalisation~\cite{Bedaque:1999ve}.

\subsubsection*{Neutron--deuteron system}

\begin{figure}[htbp]
\centering
\includegraphics[clip]{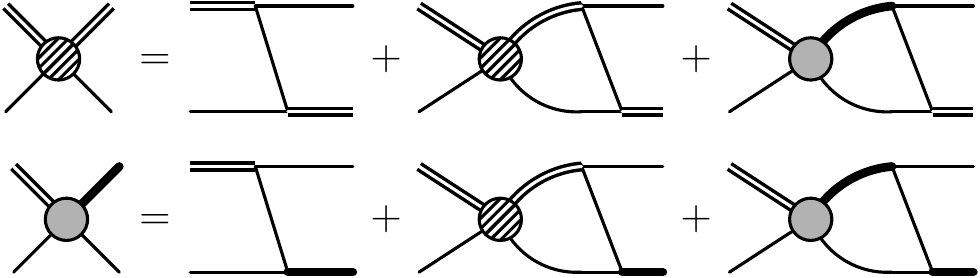}
\caption{Coupled-channel integral equation for the strong scattering amplitude
$\TS$ in the doublet channel. The diagrams involving the three-body force have
been omitted.}
\label{fig:nd-IntEq-D}
\end{figure}

As we did in the quartet channel, we look at the neutron--deuteron system first.
Fig.~\ref{fig:nd-IntEq-D} shows a diagrammatic representation of the
coupled-channel integral equation for the scattering amplitude $\TS$, of which
we only needed to consider the upper left part for the quartet-channel system.
The contribution of the three-body interaction~(\ref{eq:L-3}) is omitted here
and will be included below. After projecting onto the $n$--$d$ doublet channel
with
\begin{subequations}
\begin{align}
\label{eq:P-D-a}
 \TSda &= \frac{1}{3}(\sigma^i)^{\alpha'}_\alpha
 (\TS^{\mathrm{a},ij})^{\beta'b}_{\alpha'a}(\sigma^j)^\beta_{\beta'}
 \Big|_{\begin {subarray}{l}a=b=2\\\alpha=\beta=1\end{subarray}} \,, \\
\label{eq:P-D-b}
 \TSdb &= \frac{1}{3}(\sigma^i)^{\alpha'}_\alpha
 (\TS^{\mathrm{b},iB})^{\beta b'}_{\alpha'a}(\tau^B)^b_{b'}
 \Big|_{\begin{subarray}{l}a=b=2\\\alpha=\beta=1\end {subarray}} \,,
\end{align}
\end{subequations}
we find
\begin{subequations}
\begin{align}
\label{eq:nd-IntEq-D-a-no-H}
 \TSda &= \frac{\MN\yd^2}2\KS
  - \TSda \otimes \left[\frac{\MN\yd^2}2 D_d \KS\right]
  + \TSdb \otimes \left[\frac{3\MN\yd\yt}2 D_t \KS\right] \,, \\
\label{eq:nd-IntEq-D-b-no-H}
 \TSdb &= -\frac{3\MN\yd\yt}2\KS
  + \TSda \otimes \left[\frac{3\MN\yd\yt}2 D_d \KS\right]
  - \TSdb \otimes \left[\frac{\MN\yt^2}2 D_t \KS\right] \,.
\end{align}
\label{eq:nd-IntEq-D-no-H}
\end{subequations}
More details on the derivation and projection are again given in 
Appendix~\ref{sec:ScattEq-Details}.

\subsubsection*{Three-nucleon force}

We still need to include the contribution of the three-nucleon
interaction~(\ref{eq:L-3}) in Eqs.~(\ref{eq:nd-IntEq-D-a-no-H}) and
(\ref{eq:nd-IntEq-D-b-no-H}). A straightforward  calculation shows that this can
be achieved with the replacement~\cite{Bedaque:1999ve}
\begin{equation}
 \KS(E;k,p) \rightarrow \KS(E;k,p) + \frac{2H(\Lambda)}{\Lambda^2} \,.
\label{eq:KS-3N}
\end{equation}
It is, however, important to note that for the terms with an additional factor
of $3$ in front of $\KS$, it is cancelled in the $H(\Lambda)$-part by the
additional factor $1/3$ in~\eqref{eq:L-3}. With this, we arrive at the final
version of our integral equations,
\begin{multline}
 \begin{pmatrix}\TSda \\[\skrowspace] \TSdb\end{pmatrix}
  = \begin{pmatrix}
  g_{dd}\left(\KS+\frac{2H(\Lambda)}{\Lambda^2}\right)\\[\skrowspace]
  -g_{dt}\left(3\KS+\frac{2H(\Lambda)}{\Lambda^2}\right)
 \end{pmatrix} \\
 +\begin{pmatrix}
 -g_{dd}D_d\left(\KS+\frac{2H(\Lambda)}{\Lambda^2}\right) &
 g_{dt}D_t\left(3\KS+\frac{2H(\Lambda)}{\Lambda^2}\right) \\[\skrowspace]
 g_{dt}D_d\left(3\KS+\frac{2H(\Lambda)}{\Lambda^2}\right) &
 g_{tt}D_t\left(\KS+\frac{2H(\Lambda)}{\Lambda^2}\right)
 \end{pmatrix}
 \otimes\begin{pmatrix}\TSda\\[\skrowspace]\TSdb\end{pmatrix} \,,
\end{multline}
written in a compact matrix--vector notation. We have furthermore introduced the
abbreviations
\begin{equation}
 g_{dd} = \frac{\MN\yd^2}2 \mathtext{,}
 g_{dt} = \frac{\MN\yd\yt}2 \mathtext{,}
 g_{tt} = \frac{\MN\yt^2}2 \,.
\end{equation}
Again, the dependence on the bare coupling constants $y_d$ and $y_t$ is only
kept for notational convenience and drops out in all observables.

\subsubsection*{Proton--deuteron system}

Finally, we have all the ingredients to discuss the proton--deuteron system in
the doublet channel.

\begin{figure}[htbp]
\centering
\includegraphics[clip]{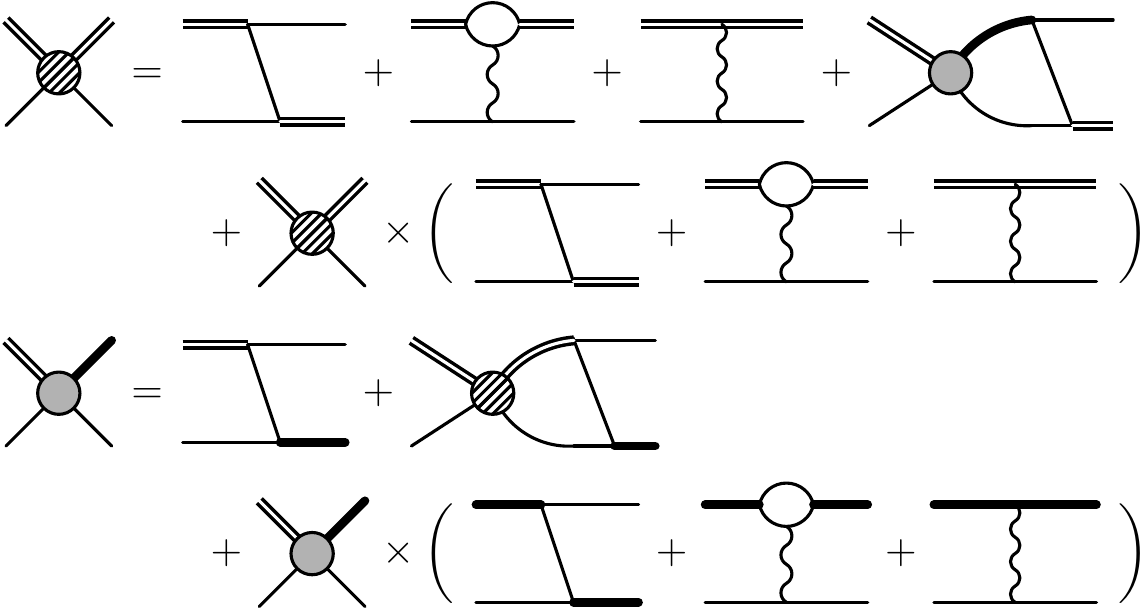}
\caption{Coupled-channel integral equation for the full (\ie\ strong + Coulomb)
scattering amplitude $\TF$ in the doublet channel. The diagrams representing the
three-nucleon force have been omitted.}
\label{fig:pd-IntEq-D}
\end{figure}

Due to the fact that the electromagnetic interaction does not couple to isospin
eigenstates we now need two different projections for the amplitude
$\mathcal{T}^{\mathrm{b}}$ with the outgoing spin-singlet dibaryon:
\begin{subequations}
\begin{align}
\label{eq:Proj-D-b1}
 \TFdb &= \frac{1}{3}(\sigma^i)^{\alpha'}_\alpha
 (\TF^{\mathrm{b},iB})^{\beta b'}_{\alpha'a}(\one\cdot\delta^{B3})^b_{b'}
 \Big|_{\begin{subarray}{l} a=b=1\\\alpha=\beta=1\end{subarray}} \,, \\
\label{eq:Proj-D-b2}
 \TFdc &= \frac{1}{3}(\sigma^i)^{\alpha'}_\alpha
 (\TF^{\mathrm{b},iB})^{\beta b'}_{\alpha'a}
 (\one\cdot\delta^{B1}+\ii\one\cdot\delta^{B2})^b_{b'}
 \Big|_{\begin{subarray}{l}a=1,\,b=2\\\alpha=\beta=1\end{subarray}} \,.
\end{align}
\label{eq:Proj-D-b1b2}
\end{subequations}
The latter corresponds to the amplitude with the outgoing spin-singlet dibaryon
in a pure $pp$-state. For the diagrams that have this component in the
intermediate state, we have to insert the propagator~\eqref{eq:Prop-t-pp-Low}
into the unprojected equations in Appendix~\ref{sec:ScattEq-Details} and find
\begin{multline}
 \begin{pmatrix}\TFda\\[\skrowspace]\TFdb\\[\skrowspace]\TFdc\end{pmatrix}
 = \begin{pmatrix}
 g_{dd}\left(\KS+\frac{2H(\Lambda)}{\Lambda^2}\right)\\[\skrowspace]
 -g_{dt}\left(\KS+\frac{2H(\Lambda)}{3\Lambda^2}\right)\\[\skrowspace]
 -g_{dt}\left(2\KS+\frac{4H(\Lambda)}{3\Lambda^2}\right)
 \end{pmatrix}
 + \begin{pmatrix}g_{dd}\KCd\\[\skrowspace]0\\[\skrowspace]0\end{pmatrix} \\
 +\begin{pmatrix}
 -g_{dd}D_d\left(\KS+\frac{2H(\Lambda)}{\Lambda^2}\right) &
 g_{dt}D_t\left(3\KS+\frac{2H(\Lambda)}{\Lambda^2}\right) & 0\\[\skrowspace]
 g_{dt}D_d\left(\KS+\frac{2H(\Lambda)}{3\Lambda^2}\right) & 
 g_{tt}D_t\left(\KS+\frac{2H(\Lambda)}{\Lambda^2}\right) & 0\\[\skrowspace]
 g_{dt}D_d\left(2\KS+\frac{4H(\Lambda)}{3\Lambda^2}\right) &
 -g_{tt}D_t\left(2\KS+\frac{4H(\Lambda)}{\Lambda^2}\right) & 0
 \end{pmatrix}
 \otimes \begin{pmatrix}
 \TFda\\[\skrowspace]\TFdb\\[\skrowspace]\TFdc\end{pmatrix} \\
 + \begin{pmatrix}
 -g_{dd}D_d\KCd & 0 & g_{dt}D_t^{pp}
 \left(3\KS+\frac{2H(\Lambda)}{\Lambda^2}\right)\\[\skrowspace]
 0 & -g_{tt}D_t\KCt & -g_{tt}D_t^{pp}
 \left(\KS+\frac{2H(\Lambda)}{\Lambda^2}\right)\\[\skrowspace]
 0 & 0 & 0
 \end{pmatrix}
 \otimes\begin{pmatrix}
 \TFda\\[\skrowspace]\TFdb\\[\skrowspace]\TFdc
 \end{pmatrix}
\label{eq:pd-IntEq-D-full}
\end{multline}
with
\begin{equation}
 D_t^{pp}(E;q)\equiv\Delta_{t,pp}\left(E-\frac{q^2}{2\MN},q\right) \,.
\label{eq:D-t-pp}
\end{equation}

The terms in~\eqref{eq:pd-IntEq-D-full} have been separated in such a way that
the sub-channels with Coulomb contributions can be easily identified. The
equation for the Coulomb scattering amplitude $\TC$ is exactly the same as in
the quartet channel.

\subsection{Higher order corrections}

The dibaryon propagators with the resummed kinetic energy insertions have an
unphysical deep bound state pole at the radius of convergence of the geometric
series. In the quartet channel the cutoff can be chosen low enough to avoid
that pole. Due to the larger cutoff needed in the doublet channel, however, we
cannot use the resummed propagators here. Instead, we use linear and quadratic
insertions of the kinetic energy operator in in the kernel of the integral
equations in order to include effective range corrections and obtain the
next-to-leading order and next-to-next-to-leading order propagators
$D_{d,t}^\NLO$ and $D_{d,t}^\NNLO$~\cite{Bedaque:2002yg}. Alternatively, we can
think of this as re-expanding the renormalised
propagators~\eqref{eq:Prop-d-High} and \eqref{eq:Prop-t-High} up to linear and
quadratic order in $\rho_{d,t}$, respectively. This still resums some higher
order effective range contributions, but removes the unphysical pole.

The question of when higher-order three-body forces enter in the doublet
channel is still under discussion. In Ref.~\cite{Bedaque:2002yg}, a subleading
three-body force was included as required by na\"{i}ve dimensional analysis. A
Lepage-plot analysis showed that its inclusion reduces the errors in the
calculation. This was supported by Ref.~\cite{Griesshammer:2005ga}, where a
corresponding logarithmic divergence at N$^2$LO, requiring a subleading
three-body force, was identified. More recently, Platter and Phillips, using the
subtractive renormalisation scheme, showed that the leading three-body force is
sufficient to achieve cutoff independence up to \NNLO~\cite{Platter:2006ev}. A
perturbative analysis recently showed that there is a new three-body parameter
already at \NLO\ if the scattering length is not fixed~\cite{Ji:2010su}. In this
work, the scattering length \emph{is} fixed and we will not include a subleading
three-body force. Assuming the counting of~\cite{Bedaque:2002yg}, our
calculation will correspond to \NNLO\ in the quartet channel and to \NLO\ in the
doublet channel. We will also perform a calculation including only the two-body
interactions to \NNLO\ in the doublet channel. 

\subsection{Numerical implementation}

The integral equations presented in the previous sections have to be solved
numerically. We do so by discretising the integrals, using Gaussian quadrature,
principal value integration to deal with the singularity of the deuteron
propagator, and appropriate transformations of the integration domain.

\begin{figure}[htbp]
\centering
\includegraphics[width=0.8\textwidth,clip]{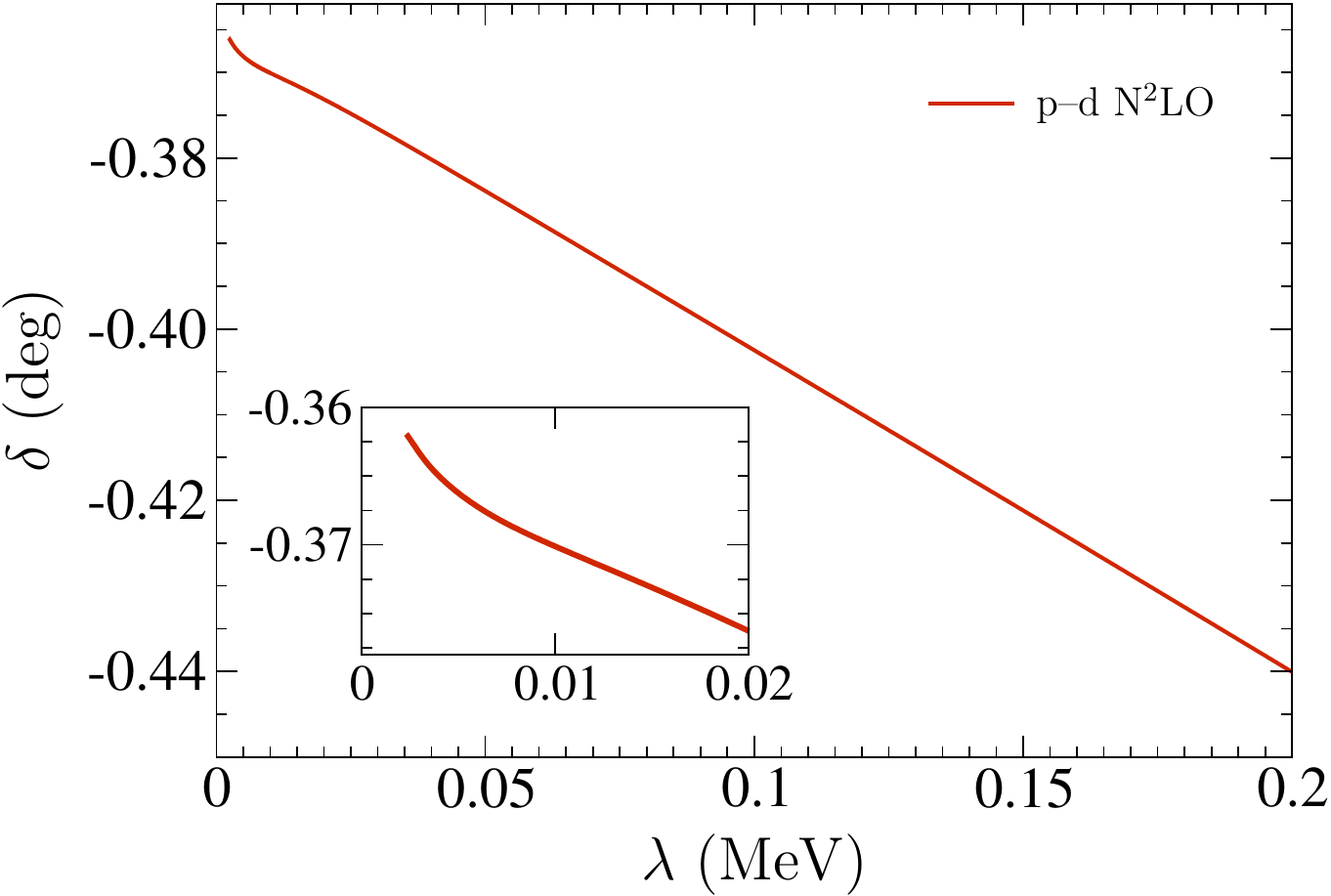}
\caption{$p$--$d$ quartet channel S-wave scattering phase shift
at \NNLO\ for center-of-mass momentum $k=5~\MeV$ and cutoff $\Lambda=140~\MeV$
as a function of the regulating photon mass $\lambda$.}
\label{fig:Phase-l-Q}
\end{figure}

The latter are especially important to deal with the numerical difficulties
caused by the Coulomb photon propagators. Even though we have regulated the
singularity with the artificial photon mass $\lambda$, the latter has to be kept
small, which then yields strongly peaked functions. It turns out that
the Coulomb peaks in the inhomogeneous parts of Eqs.~\eqref{eq:pd-IntEq-Q-full},
\eqref{eq:pd-IntEq-Q-c} and \eqref{eq:pd-IntEq-D-full} are the major
numerical problem. We solve it by concentrating the quadrature points around
this peak. Together with always putting half of the quadrature points into the
low-momentum region (the interval from zero to the peak position), we are able
to (linearly) extrapolate our results for the scattering phase shifts back to
the physical value $\lambda=0$ (screening limit). As a typical example, we show
the quartet-channel phase shift at $k=5~\MeV$ as a function of $\lambda$ in
Fig.~\ref{fig:Phase-l-Q}. The linear dependence is clearly visible, with
deviations only for very small $\lambda$. In the doublet channel, the
qualitative behaviour is the same, only the curvature becomes visible already
for somewhat larger photon masses. This can be understood by noting that the
absolute value of the phase shift is smaller in the doublet channel, especially
for low center-of-mass momenta $k$ (\cf~Sec.~\ref{sec:Scatt-Results}). We
settled to use the intervals $0.1\leq\lambda\leq0.15$ and
$0.4\leq\lambda\leq0.6$ for the extrapolations in the quartet and doublet
channel, respectively. We note that the error introduced by the extrapolation to
$\lambda=0$ can generally be neglected compared to the theoretical error from
the EFT expansion discussed below.

We have used the experimental input parameters shown in Tab.~\ref{tab:Params} in
the numerical calculation.

\begin{table}[htbp]
 \begin{tabular}{cc|cc}
  Parameter & Value & Parameter & Value \\
  \hline
  $\gamd$ & $45.701~\MeV$ \cite{vanderLeun:1982aa} & $\rd$
  & $1.765~\fm$ \cite{deSwart:1995ui} \\
  $a_t$ & $-23.714~\fm$ \cite{Beane:2000fx} & $\rnt$
  & $2.73~\fm$ \cite{Beane:2000fx} \\
  $a_C$ & $-7.8063~\fm$ \cite{Bergervoet:1988zz}& $\rnC$
  & $2.794~\fm$ \cite{Bergervoet:1988zz}\\
 \end{tabular}
\caption{Parameters used in the numerical calculation, $\gamd=\sqrt{\MN\EB^d}$.}
\label{tab:Params}
\end{table}

\section{Scattering results}
\label{sec:Scatt-Results}

\subsection{Quartet channel}

\begin{figure}[htbp]
\centering
\includegraphics[width=0.8\textwidth,clip]{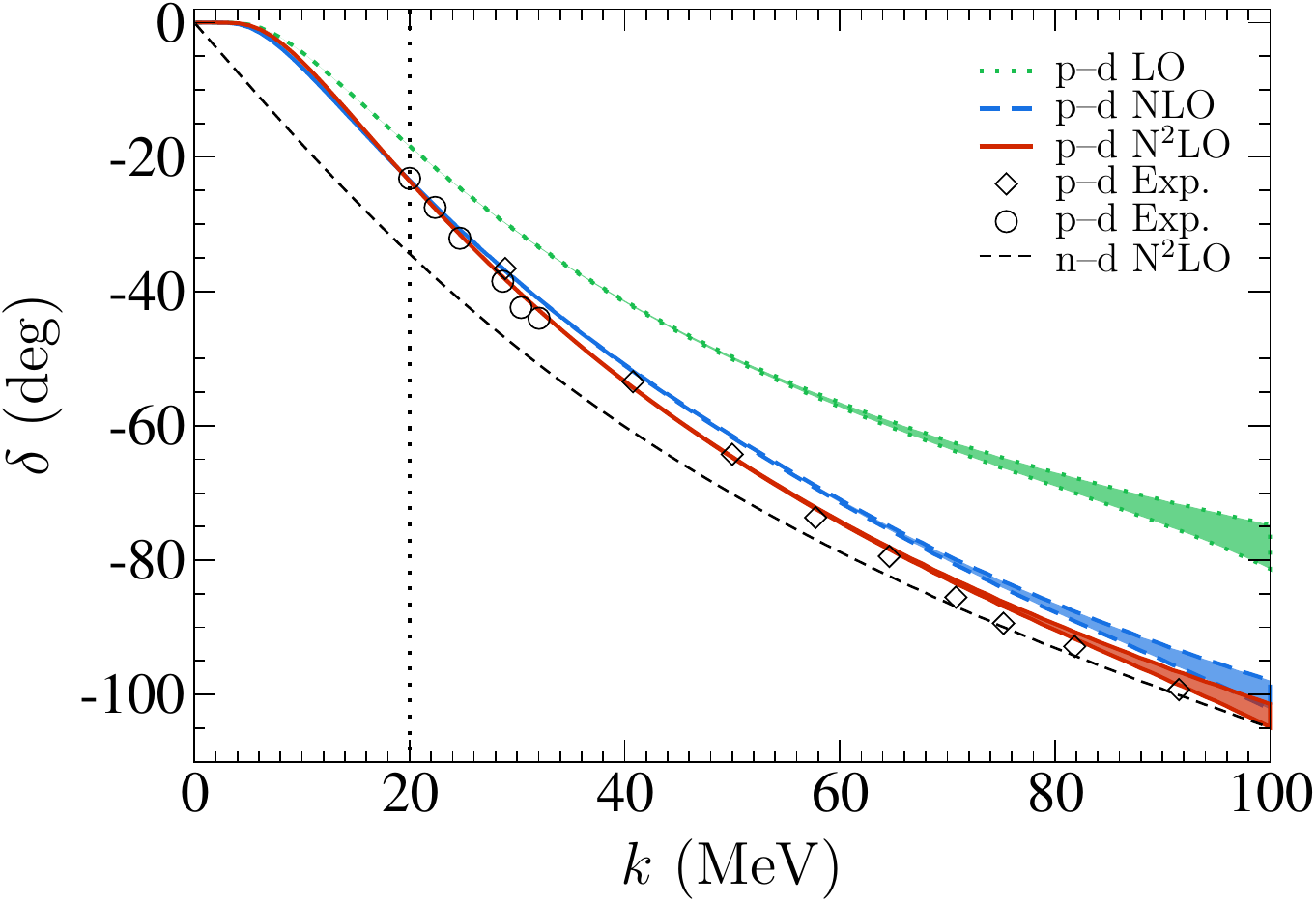}
\caption{$N$--$d$ quartet channel S-wave scattering phase shifts
as functions of the center-of-mass momentum $k$. Error bands generated by cutoff
variation from 120 to 160~\MeV. Experimental $p$--$d$ phase shift data taken
from \cite{Arvieux:1973ab} (diamonds) and \cite{Christian:1953ab} (circles).}
\label{fig:Phase-Nd-Q}
\end{figure}

In Fig.~\ref{fig:Phase-Nd-Q} we show the phase shift results for both
neutron--deuteron and proton--deuteron scattering as functions of the
center-of-mass momentum $k$. The error bands are generated by varying the cutoff
within a range of $120$ to $160~\MeV$. Since the cutoff variation is small, we
conclude that the calculation is well converged at these cutoffs. For
$\Lambda\gsim 200~\MeV$ some numerical artifacts show up from integrating over
the unphysical second pole in the full deuteron propagator. For the $n$--$d$
curve in Fig.~\ref{fig:Phase-Nd-Q} we have used $\Lambda=140~\MeV$. The fact
that the bands do not overlap is no point of concern since they only give a
lower bound on the error of the calculation. From the expansion parameter
$\gamd\rd\approx1/3$ of the EFT, the error can be estimated as 30\%, 10\%, and
3\% at \LO, \NLO, and \NNLO, respectively. Thus, at \LO, the 30\% error from the
expansion parameter clearly dominates. At \NLO\ and \NNLO, however, the band
from the cutoff variation gives a reasonable estimate of the total error in the
calculation. 

The \NNLO~result to the right of the dotted line at $k=20~\MeV$ agrees nicely
with the results presented in~\cite{Rupak:2001ci} and also with the experimental
data included in the plot. At this point we remark, however, that for
$k\gsim20~\MeV$ the Coulomb parameter $\alpha\MN/k$ is of order $1/3$, which
means that in this regime the non-perturbative treatment of Coulomb effects
might not even be necessary. More important are hence the $p$--$d$ results for
small momenta ($k < 20$ MeV) to the left of the dotted line, which we could
obtain thanks to our optimised numerical procedure. It would of course be good
to have some data points in this region to test our prediction.

\subsection{Doublet channel}

\begin{figure}[phtb]
\centering
\includegraphics[width=0.8\textwidth,clip]{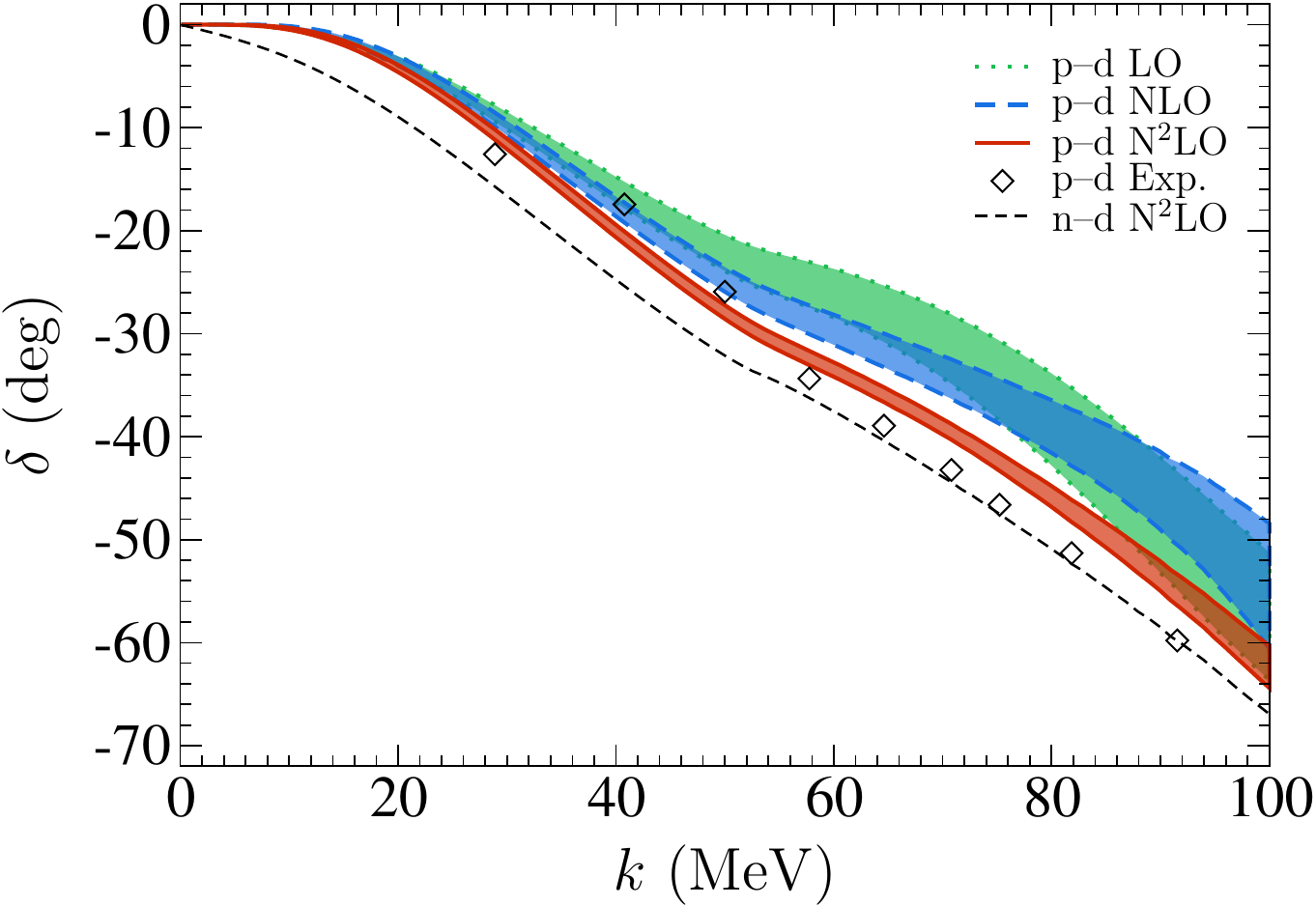}
\caption{$N$--$d$ doublet channel S-wave scattering phase shifts
as functions of the center-of-mass momentum $k$. Error bands generated by cutoff
variation from 200 to 600~\MeV. Experimental $p$--$d$ phase shift data taken
from \cite{Arvieux:1973ab}.}
\label{fig:Phase-Nd-D}
\end{figure}

The doublet-channel results for the $p$--$d$ scattering phase shifts as
functions of the center-of-mass momentum $k$ are shown in
Fig.~\ref{fig:Phase-Nd-D}. As in the quartet channel, the $p$--$d$ curve lies
above the $n$--$d$ curve and agrees quite well with the experimental data. A
more quantitative comparison is, unfortunately, not possible since there are no
errors given for the data points. The error bands are generated by varying the
cutoff within a natural range of $200$ to $600~\MeV$ (\ie\ a few times the pion
mass). Assuming the power counting of~\cite{Bedaque:2002yg}, our 
\NNLO~calculation is incomplete since the subleading three-body force is not
included. A full calculation, however, is beyond the scope of the paper since
gauging the subleading three-body force creates new three-body contributions to
the photon coupling. Our partial \NNLO~result is stable to within about ten
percent under the cutoff variation, which is consistent with a 7--15\% error
estimate based on the neglected Coulomb diagrams (see
Sec.~\ref{sec:Power-Counting} and~\cite{Rupak:2001ci}).

Figure~\ref{fig:Phase-Nd-D} furthermore shows how the results improve from order
to order. The stability of the partial \NNLO~result with respect to variation of
the cutoff suggests that the scattering is relatively insensitive to the 
subleading three-body interaction. At higher energies, however, there is some
room for such a contribution as a our partial result consistently lies two to
four degrees above the data.

We observe that the shift from \LO\ to \NLO\ is of the same order of magnitude
as the shift from \NLO\ to \NNLO. This behaviour is typical for effective range
corrections in the doublet channel~\cite{Platter:2006ad}. The smallness of the
\NLO~corrections can be understood as a cancellation between two different 
contributions to this correction. The two contributions are proportional to
$\kappa \rd$ and $\gamd\rd$, respectively, where $\kappa$ is the typical
momentum scale of the process. Furthermore, it is known that at \LO~observables
are often described better than expected from the power counting once the exact
pole position of the two-body propagator is reproduced~\cite{Braaten:2004rn}. As
a consequence, the shifts in observables from \LO\ to \NLO\ can be small and of
a size comparable to the corresponding shifts from \NLO\ to \NNLO.

\section{$^3$He bound state properties}
\label{sec:He-3}

Since the power counting we used for the Coulomb contributions is not valid for
the regime of typical bound state energies, we cannot simply use the above
equations to calculate $^3$He. There are various strategies to proceed. We
could extend the power counting and include additional Coulomb diagrams in our
equations. Alternatively, we could use an analytic expression for the full
off-shell Coulomb amplitude as it is done in Ref.~\cite{Ando:2010wq} for an
\LO~calculation. Here, we choose the much simpler approach of calculating the
Coulomb energy shift for $^3$He in first order perturbation theory as the
expectation value of the Coulomb interaction between proton and deuteron using
trinucleon wave functions in the isospin limit.

\subsection{Trinucleon wave functions}

In order to obtain the trinucleon wave functions we need to solve the
homogeneous coupled-channel equation
\begin{equation}
 \vec{\BS} = (\hat{K}\hat{D})\otimes\vec{\BS}
\label{eq:BS-IntEq}
\end{equation}
with $\vec{\BS}\equiv\left(\BSda,\BSdb,\BSdc\right)^T$, $\hat D =
\diag(D_d,D_t,D_t)$, and
\begin{equation}
 \hat{K}\equiv\begin{pmatrix}
 -g_{dd}\left(\KS+\frac{2H(\Lambda)}{\Lambda^2}\right) &
 g_{dt}\left(3\KS+\frac{2H(\Lambda)}{\Lambda^2}\right) &
 g_{dt}\left(3\KS+\frac{2H(\Lambda)}{\Lambda^2}\right)\\[\skrowspace]
 g_{dt}\left(\KS+\frac{2H(\Lambda)}{3\Lambda^2}\right) &
 g_{tt}\left(\KS+\frac{2H(\Lambda)}{\Lambda^2}\right) &
 -g_{tt}\left(\KS+\frac{2H(\Lambda)}{\Lambda^2}\right)\\[\skrowspace]
 g_{dt}\left(2\KS+\frac{4H(\Lambda)}{3\Lambda^2}\right) &
 -g_{tt}\left(2\KS+\frac{4H(\Lambda)}{\Lambda^2}\right) & 0
 \end{pmatrix} \,.
\end{equation}
It is obtained by applying the projections~\eqref{eq:Proj-D-b1b2} to the raw
equation without Coulomb contributions~\eqref{eq:nd-IntEq-D-raw}. The reason to
separate the wave function in this way is that the part with the dibaryon leg in
the $pp$-channel does not contribute to a perturbative calculation of the energy
shift since in that case the third nucleon in the system necessarily is a
neutron. The energy in the equation is set to the experimental triton binding
energy,
\begin{equation}
 {-\EB}^\mathrm{\!\!^3\mathrm{H}} = -8.48~\MeV \,,
\label{eq:E-Triton}
\end{equation}
and the existence of solution is ensured by adjusting the three-nucleon force
$H(\Lambda)$ appropriately, as it was already done to renormalise the scattering
equations.

Having obtained the wave functions as solutions of~\eqref{eq:BS-IntEq}, we still
need to normalise them properly. Since the EFT generates an energy-dependent
interaction in the three-body system, this is done by demanding that
\begin{equation}
 \left(\hat D\vec{\BS}\right)^T
 \otimes\frac\dd{\dd E}\left(\hat I-\hat K\right)
 \Big|_{E=-\EB^\mathrm{\,^3\mathrm{H}}}
 \otimes\left(\hat D\vec{\BS}\right) = 1 \,,
\label{eq:Triton-WF-norm}
\end{equation}
where $\hat I = \diag(I_d,I_t,I_t)$ with
\begin{equation}
 I_{d,t}(E,q,q') = \frac{2\pi^2}{q^2}{\delta(q-q')}D_{d,t}(E;q)^{-1} \,.
\end{equation}
A short derivation of this normalisation condition can be found in
Appendix~\ref{sec:Triton-Norm}.

\subsection{Perturbative $^3$He energy shift}

With the normalised trinucleon wave functions we can obtain the Coulomb-induced
energy shift in first order perturbation theory and find
\begin{equation}
 \Delta E = \left(\hat D\vec{\BS}\right)^T\otimes\diag(V_C,V_C,0)
 \otimes\left(\hat D\vec{\BS}\right)
\end{equation}
with the S-wave projected Coulomb potential
\begin{equation}
 V_C(E;q,q') = -\frac{4\pi\alpha}{2qq'}\;
 Q\left(-\frac{q^2+q'^2+\lambda^2}{2qq'}\right)
\label{eq:V-C}
\end{equation}
in momentum space. Our prediction for the $^3$He binding energy is then given by
\begin{equation}
 {-\EB}^{\!\!^3\mathrm{He}} = {-\EB}^\mathrm{\!\!^3\mathrm{H}}+\Delta E \,.
\end{equation}

\subsection{Results}

\begin{figure}[htbp]
\centering
\includegraphics[width=0.8\textwidth,clip]{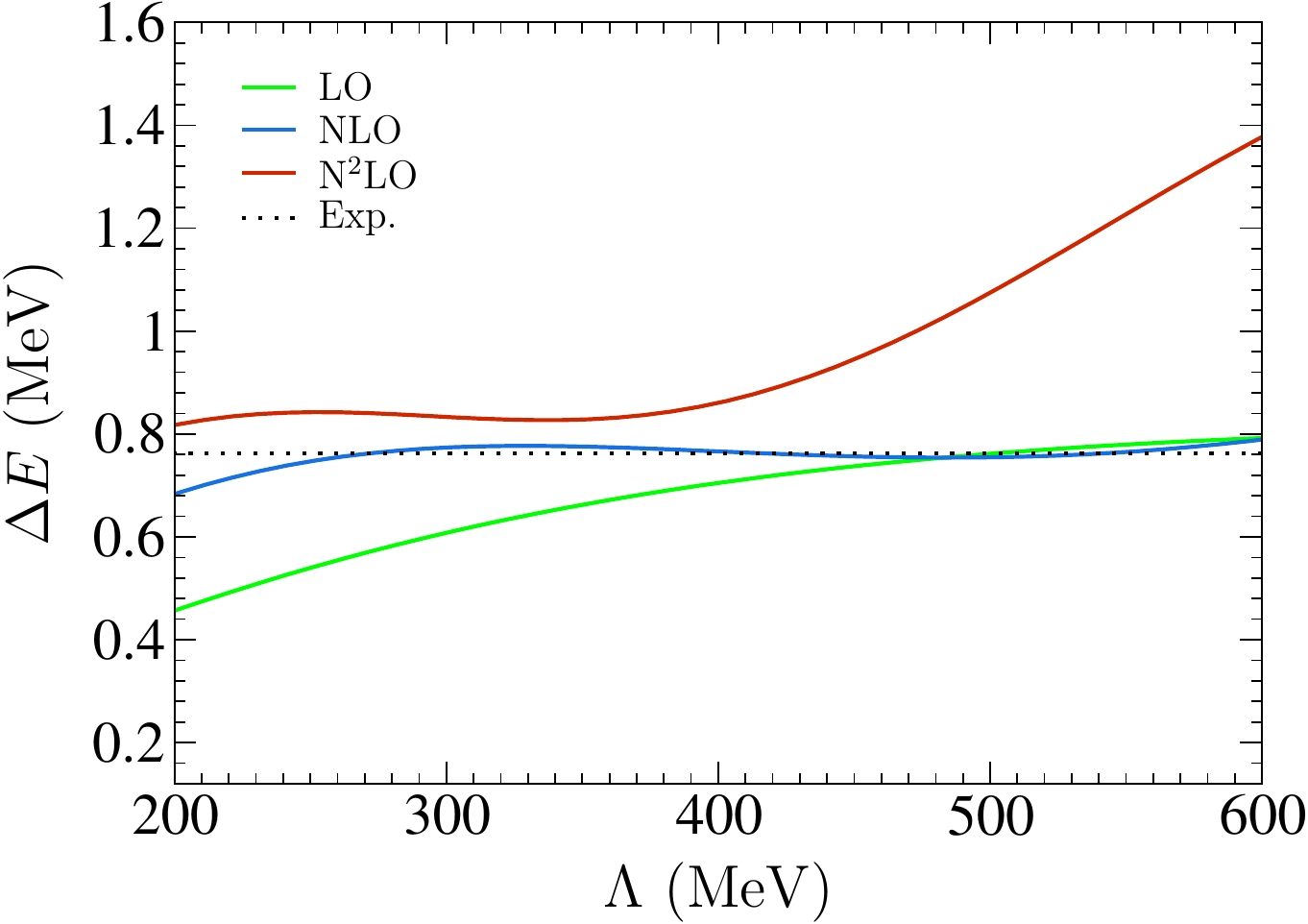}
\caption{Perturbative prediction for the $^3$He--$^3$H binding
energy difference in dependence of the cutoff $\Lambda$. Bottom curve:
\LO~result. Middle curve: \NLO~result. Top curve: \NNLO~result.}
\label{fig:En-WF}
\end{figure}

The results are shown in Fig.~\ref{fig:En-WF}. It is remarkable that the
\NLO~result, which should be accurate to about 10\%, agrees very well with the
experimental value
\begin{equation}
 \Delta E_\mathrm{exp} = 0.7629~\MeV
\end{equation}
over a large cutoff range. In our partial \NNLO~calculation, the results are
still quite stable against cutoff variations within $200$ to about $400~\MeV$,
but our value lies about $0.1~\MeV$ above the experimental value. This shift
again leaves room for a natural-sized contribution of the subleading three-body
we have not included.

Brandenburger, Coon and Sauer have determined the $^3$He--$^3$H binding energy
difference in a largely model-independent way using experimental charge form
factors \cite{Coon78}. They found that the Coulomb contribution is about 10\%
below the experimental value for the total energy difference. Within the
expected error of 10\%, our \NLO~result is consistent with this.
Ando and Birse have carried out a non-perturbative calculation to leading order
in the pionless EFT including the full off-shell $\mathcal{T}$-matrix for the
Coulomb interaction and found $\Delta E = 0.82~\MeV$~\cite{Ando:2010wq}. Their
calculation included isospin breaking effects in the nucleon--nucleon scattering
lengths. Kirscher \textit{et al}. found the smaller value $\Delta E = 0.66 \pm
0.03~\MeV$ in an \NLO~calculation using the resonating group model to solve the
pionless EFT with a charge-independent value of the spin-singlet scattering
length and non-perturbative Coulomb interactions~\cite{Kirscher:2009aj}.

\begin{figure}[htbp]
\centering
\includegraphics[width=0.8\textwidth,clip]{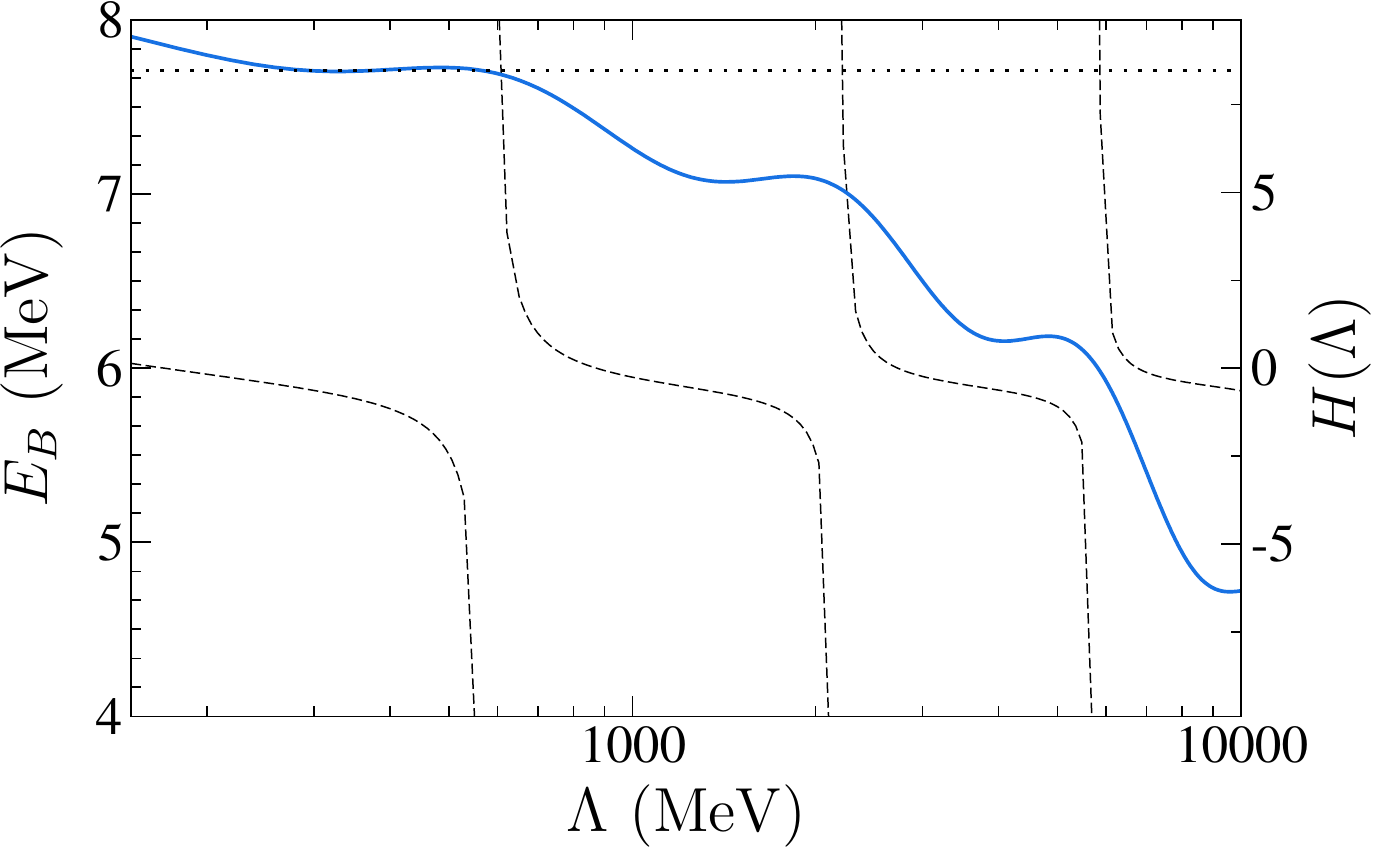}
\caption{Perturbative prediction for the $^3$He binding energy
together with the three-nucleon force; \NLO~results. Solid curve: $^3$He binding
energy prediction. Dashed curve: three-nucleon force. Dotted line: experimental
value for the $^3$He binding energy.}
\label{fig:H-En-NLO}
\end{figure}

The increase in our \NNLO~result for $\Lambda\gsim400~\MeV$ that is seen in
Fig.~\ref{fig:En-WF} also occurs at \LO\ and \NLO, but for larger cutoffs. In
Fig.~\ref{fig:H-En-NLO} we show the \NLO~prediction for
$\EB^{^3\mathrm{He}}$ together with the three-nucleon force obtained from
fitting the triton binding energy. It is obvious that a drop in the binding
energy prediction occurs whenever the three-nucleon force has gone through a
pole. We interpret this as an artifact of the theory which is related to the
Efimov effect. For cutoffs beyond the position of the first pole in
$H(\Lambda)$, the triton is not the true ground state of the system anymore;
after each pole transition a new (unphysical) deep bound state emerges.

These unphysical deep states lead to additional nodes in the triton wave
function at short distances which affect our perturbative results. This suggests
that an additional short-distance counterterm is required to cancel these
contributions if one wants to go to cutoffs much larger than the pion mass.

\section{Summary and outlook}

In this paper, we have investigated S-wave proton--deuteron scattering in
pionless effective field theory. In the quartet channel, we have calculated the
elastic scattering phase shift up to \NNLO\ using the power counting for Coulomb
contributions suggested by Rupak and Kong~\cite{Rupak:2001ci}. The Coulomb
effects are included at \NLO~accuracy in our calculation. Using an optimised
integration mesh we were able to extend their calculation into the threshold
region were the Coulomb interaction becomes highly non-perturbative. We found
good agreement both with available phase shift analyses and with the results of
Rupak and Kong at momenta $k\geq 20~\MeV$.

Moreover, we extended the power counting to the doublet channel and performed
a complete calculation of the phase shifts to \NLO\ in agreement with the
available phase shift data. We also carried out a partial \NNLO~calculation
that neglected the contribution of the subleading three-body force entering at
this order. The results of this calculation are stable under variations of the
cutoff. Furthermore, there is good agreement with the phase shift data at
low momenta and room for a small contribution of the neglected three-body force
at larger momenta. Overall, however, the doublet channel phase shifts are only
weakly sensitive to the subleading three-body force entering at \NNLO.

Although we were mainly interested in $p$--$d$ scattering, we have also
calculated the Coulomb contribution to the $^3$He--$^3$H binding energy
difference $\Delta E$. This observable has previously been calculated in the
pionless theory by treating the Coulomb interaction
non-perturbatively~\cite{Ando:2010wq,Kirscher:2009aj}. Here, we treat the
Coulomb potential between proton and deuteron in first order perturbation theory
using trinucleon wave functions. Higher order corrections to this quantity are
expected to be small. Our \NLO~result is in reasonable agreement with the
experimental value and other evaluations. We find $\Delta E$ to be more
sensitive to the subleading three-body force. The partial \NNLO~result is about
10\% too large, thus leaving room for a contribution from the omitted three-body
force. We also observe steps in the calculated value of $\Delta E$ as the cutoff
is increased beyond its natural range. Whenever the leading three-nucleon force
has gone through a pole, a drop in the calculated binding energy occurs. We
interpret this as an artifact of the theory related to the Efimov effect. At
higher cutoffs, spurious deep three-body bound states appear and the triton is
not the true ground state anymore. It appears that an additional short-distance 
counterterm is required to cancel these contributions if one wants to go to
cutoffs much larger than the pion mass. A further study of this issue would be
interesting.

In the future, a full \NNLO~calculation including the subleading three-body
force and the electromagnetic interaction terms generated from gauging
its momentum dependence should be carried out. Such an accuracy will, \eg,
be required for high-precision calculations of low-energy astrophysical
processes in pionless effective field theory and the effective field theory for
halo nuclei.

\begin{acknowledgments}
We thank A.~Rusetsky, B.~Metsch, M.~Hoferichter and P.~Hagen for discussions and
S.-I.~Ando, S.~Coon, D.~R.~Phillips and U.~van~Kolck for comments on the
manuscript. This research was supported in part by the DFG through SFB/TR~16
``Subnuclear structure of matter'' and the BMBF under contract No.~06BN9006.
S.K. was supported by the ``Studien\-stiftung des deutschen Volkes'' and by the
Bonn-Cologne Graduate School of Physics and Astronomy.
\end{acknowledgments}

\appendix

\section{Scattering equation details}
\label{sec:ScattEq-Details}

\subsection{Quartet channel}

Using the Feynman rules that follow from the Lagrangian~\eqref{eq:L-Nd} and
inserting appropriate symmetry factors, we find
\begin{multline}
 (\ii\TS^{ij})^{\beta b}_{\alpha a}(E;\vk,\vp)
 = -\frac{\ii\MN\yd^2}{2}\cdot(\sigma^j\sigma^i)^\beta_\alpha\delta^b_a
 \cdot\frac{1}{\vk^2+\vk\cdot\vp+\vp^2-\MN E-\ii\eps} \\
 +\int\ddq\,\Delta_d\left(E-\frac{\vq^2}{2\MN},\vq\right)
 \cdot(\ii\TS^{ik})^{\gamma c}_{\alpha a}(E;\vk,\vq) \\
 \times\frac{\MN\yd^2}{2}\frac{(\sigma^j\sigma^k)^\beta_\gamma\delta^b_c}
 {\vq^2+\vq\cdot\vp+\vp^2-\MN E-\ii\eps}
\label{eq:nd-IntEq-Q-raw}
\end{multline}
for the $n$--$d$ scattering equation depicted in Fig.~\ref{fig:nd-IntEq-Q}. In
the same way we get
\begin{multline}
 (\ii\TF^{ij})^{\beta b}_{\alpha a}(E;\vk,\vp)
  = -\frac{\ii\MN\yd^2}{2}\cdot(\sigma^j\sigma^i)^\beta_\alpha\delta^b_a
 \cdot\frac{1}{\vk^2+\vk\cdot\vp+\vp^2-\MN E-\ii\eps} \\
 -\ii\alpha\MN^2\yd^2\cdot\delta^{ij}\delta^\beta_\alpha
 \left(\frac{\one+\tau_3}{2}\right)^b_a
 \left[\frac{\mathcal{I}_\mathrm{bubble}(E;\vk,\vp)}{(\vk-\vp)^{2}+\lambda^2}
 -\frac{\rd}{2}\frac{1}{(\vk-\vp)^2+\lambda^2}\right] \\
 +\int\ddq\,\Delta_d\left(E-\frac{\vq^2}{2\MN},\vq\right)
 \cdot(\ii\TF^{ik})^{\gamma c}_{\alpha a}(E;\vk,\vq)
 \cdot\left\{\frac{\MN\yd^2}{2}\frac{(\sigma^j\sigma^k)^\beta_\gamma\delta^b_c}
 {\vq^2+\vq\cdot\vp+\vp^2-\MN E-\ii\eps}\right. \\
 +\left.\,\alpha\MN^2\yd^2\cdot\delta^{kj}\delta^\beta_\gamma
 \left(\frac{\one+\tau_3}{2}\right)^b_c
 \left[\frac{\mathcal{I}_\mathrm{bubble}(E;\vq,\vp)}{(\vq-\vp)^{2}+\lambda^2}
 -\frac{\rd}{2}\frac{1}{(\vq-\vp)^2+\lambda^2}\right]\right\}
\end{multline}
for the full $p$--$d$ scattering equation shown in Fig.~\ref{fig:pd-IntEq-Q},
and
\begin{multline}
 (\ii\TC^{ij})^{\beta b}_{\alpha a}(E;\vk,\vp)
 = -\ii\alpha\MN^2\yd^2\cdot\delta^{ij}\delta^\beta_\alpha
 \left(\frac{\one+\tau_3}{2}\right)^b_a
 \Biggl[\frac{\mathcal{I}_\mathrm{bubble}(E;\vk,\vp)}{(\vk-\vp)^{2}+\lambda^2}
 -\frac{\rd}{2}\frac{1}{(\vk-\vp)^2+\lambda^2}\Biggr] \\
 +\int\ddq\,\Delta_d\left(E-\frac{\vq^2}{2\MN},\vq\right)
 \cdot(\ii\TC^{ik})^{\gamma c}_{\alpha a}(E;\vk,\vq) \\
 \times\alpha\MN^2\yd^2\cdot\delta^{jk}\delta^\beta_\gamma
 \left(\frac{\one+\tau_3}{2}\right)^b_c
 \left[\frac{\mathcal{I}_\mathrm{bubble}(E;\vq,\vp)}
 {(\vq-\vp)^{2}+\lambda^2}-\frac{\rd}{2}\frac{1}{(\vq-\vp)^2+\lambda^2}\right]
\end{multline}
for the pure Coulomb scattering equation (Fig.~\ref{fig:pd-IntEq-Coulomb}),
where
\begin{equation}
 \mathcal{I}_\mathrm{bubble}(E;\vk,\vp)
 = \frac{\arctan\left(\frac{2\vp^2-\vk^2-\vk\cdot\vp}
 {\sqrt{3\vk^2-4\MN E-\ii\eps}\sqrt{(\vk-\vp)^2}}\right)
 +\arctan\left(\frac{2\vk^2-\vp^2-\vk\cdot\vp }
 {\sqrt{3\vp^2-4\MN E-\ii\eps}\sqrt{(\vk-\vp)^2}}\right)}
 {\sqrt{(\vk-\vp)^2}}
\label{eq:I-bubble-pd}
\end{equation}
corresponds to the loop integral in the diagram with the photon attached to a
nucleon bubble. The expression looks quite complicated, but it can be
simplified. The dominant terms of $\mathcal{I}_\mathrm{bubble}$ are those with
$\vp^2\approx\vk^2$ and $\vp^2\approx\vq^2$, respectively, due to the prefactors
of $1/(\vk-\vp)^2$ and $1/(\vq-\vp)^2$. In the latter case we can furthermore
assume that $\vq^2\approx\vk^2$ because of the pole at this position in the
propagator. Furthermore inserting the total center-of-mass energy $E=3k^2/(4\MN)
-\gamd^2/\MN$, we get
\begin{subequations}
\begin{equation}
 \frac{\mathcal{I}_\mathrm{bubble}(E;\vk,\vp)}{(\vk-\vp)^{2}+\lambda^2}
 \approx \frac{1}{2\left|\gamd\right|}\frac{1}{(\vk-\vp)^2+\lambda^2}
\label{eq:Bubble-approx-1}
\end{equation}
and
\begin{equation}
 \Delta_d\left(E-\frac{\vq^2}{2\MN},\vq\right)
 \cdot\frac{\mathcal{I}_\mathrm{bubble}(E;\vq,\vp)}{(\vq-\vp)^{2}+\lambda^2}
 \approx \Delta_d\left(E-\frac{\vq^2}{2\MN},\vq\right)
 \cdot\frac{1}{2\left|\gamd\right|}\frac{1}{(\vq-\vp)^2+\lambda^2} \,,
\end{equation}
\label{eq:Bubble-approx}
\end{subequations}
where we have used~\eqref{eq:I-bubble-pd} and the expansion
$\arctan(x)=x+\OO(x^3)$. The same simplifications, which effectively amount to
keeping only loop contributions with $q\sim p$, are used in~\cite{Rupak:2001ci}
and appear to be well supported by comparing the results with experimental data
(see Sec.~\ref{sec:Scatt-Results}).

\subsection{Doublet channel}

In the doublet channel we find
\begin{subequations}
\begin{equation}
\begin{split}
 (\ii\TS^{\mathrm{a},ij})^{\beta b}_{\alpha a}(E;\vk,\vp) =
 &-\frac{\ii\MN\yd^2}{2} \cdot(\sigma^j\sigma^i)^\beta_\alpha\delta^b_a
 \cdot\frac{1}{\vk^2+\vk\cdot\vp+\vp^2-\MN E-\ii\eps} \\
 &+\int\ddq\,\Delta_d\left(E-\frac{\vq^2}{2\MN},\vq\right)
 \cdot(\ii\TS^{\mathrm{a},ik})^{\gamma c}_{\alpha a}(E;\vk,\vq) \\
 &\hspace{10em}\times\frac{\MN\yd^2}{2}
 \frac{(\sigma^j\sigma^k)^\beta_\gamma\delta^b_c}
 {\vq^2+\vq\cdot\vp+\vp^2-\MN E-\ii\eps} \\
 &+\int\ddq\,\Delta_t\left(E-\frac{\vq^2}{2\MN},\vq\right)
 \cdot(\ii\TS^{\mathrm{b},iC})^{\gamma c}_{\alpha a}(E;\vk,\vq) \\
 &\hspace{10em}\times\frac{\MN\yd\yt}{2}
 \frac{(\sigma^j)^\beta_\gamma(\tau^C)^b_c }
 {\vq^2+\vq\cdot\vp+\vp^2-\MN E-\ii\eps}
\end{split}
\label{eq:nd-IntEq-D-a-raw}
\end{equation}
\begin{equation}
\begin{split}
 (\ii\TS^{\mathrm{b},iB})^{\beta b}_{\alpha a}(E;\vk,\vp) =
 &-\frac{\ii\MN\yd\yt}{2}\cdot(\sigma^i)^\beta_\alpha(\tau^B)^b_a
 \cdot\frac{1}{\vk^2+\vk\cdot\vp+\vp^2-\MN E-\ii\eps} \\
 &+\int\ddq\,\Delta_d\left(E-\frac{\vq^2}{2\MN},\vq\right)
 \cdot(\ii\TS^{\mathrm{a},ik})^{\gamma c}_{\alpha a}(E;\vk,\vq) \\
 &\hspace{10em}\times\frac{\MN\yd\yt}{2}
 \frac{(\sigma^k)^\beta_\gamma(\tau^B)^b_c }
 {\vq^2+\vq\cdot\vp+\vp^2-\MN E-\ii\eps} \\
 &+\int\ddq\,\Delta_t\left(E-\frac{\vq^2}{2\MN},\vq\right)
 \cdot(\ii\TS^{\mathrm{b},iC})^{\gamma c}_{\alpha a}(E;\vk,\vq)\\
 &\hspace{10em}\times\frac{\MN\yt^2}{2}
 \frac{\delta^\beta_\gamma(\tau^B\tau^C)^b_c}
 {\vq^2+\vq\cdot\vp+\vp^2-\MN E-\ii\eps}
\end{split}
\label{eq:nd-IntEq-D-b-raw}
\end{equation}
\label{eq:nd-IntEq-D-raw}
\end{subequations}
for the coupled-channel $n$--$d$ equation shown in Fig.~\ref{fig:nd-IntEq-D},
and analogously we have
\begin{subequations}
\begin{multline}
 (\ii\TF^{\mathrm{a},ij})^{\beta b}_{\alpha a}(E;\vk,\vp)
 = -\frac{\ii\MN\yd^2}{2}\cdot(\sigma^j\sigma^i)^\beta_\alpha\delta^b_a
 \cdot\frac{1}{\vk^2+\vk\cdot\vp+\vp^2-\MN E-\ii\eps} \\
 -\ii\alpha\MN^2\yd^2\cdot\delta^{ij}\delta^\beta_\alpha
 \left(\frac{\one+\tau_3}{2}\right)^b_a
 \Biggl[\frac{\mathcal{I}_\mathrm{bubble}(E;\vk,\vp)}{(\vk-\vp)^{2}+\lambda^2}
 -\frac{\rd}{2}\frac{1}{(\vk-\vp)^2+\lambda^2}\Biggr] \\
 +\int\ddq\,\Delta_d\left(E-\frac{\vq^2}{2\MN},\vq\right)
 \cdot(\ii\TF^{\mathrm{a},ik})^{\gamma c}_{\alpha a}(E;\vk,\vq)
 \cdot\Biggl\{\frac{\MN\yd^2}{2}\frac{(\sigma^j\sigma^k)^\beta_\gamma\delta^b_c}
 {\vq^2+\vq\cdot\vp+\vp^2-\MN E-\ii\eps} \\
 +\alpha\MN^2\yd^2\cdot\delta^{kj}\delta^\beta_\gamma
 \left(\frac{\one+\tau_3}{2}\right)^b_c
 \left[\frac{\mathcal{I}_\mathrm{bubble}(E;\vq,\vp)}{(\vq-\vp)^{2}+\lambda^2}
 -\frac{\rd}{2}\frac{1}{(\vq-\vp)^2+\lambda^2}\right]\Biggr\} \\
 +\int\ddq\,\Delta_t\left(E-\frac{\vq^2}{2\MN},\vq\right)
 \cdot(\ii\TF^{\mathrm{b},iC})^{\gamma c}_{\alpha a}(E;\vk,\vq)
 \cdot\frac{\MN\yd\yt}{2}\cdot\frac{(\sigma^j)^\beta_\gamma(\tau^C)^b_c}
 {\vq^2+\vq\cdot\vp+\vp^2-\MN E-\ii\eps}
\label{eq:pd-IntEq-D-a-raw}
\end{multline}
\begin{multline}
 (\ii\TF^{\mathrm{b},iB})^{\beta b}_{\alpha a}(E;\vk,\vp)
 = -\frac{\ii\MN\yd\yt}{2}\cdot(\sigma^i)^\beta_\alpha(\tau^B)^b_a
 \cdot\frac{1}{\vk^2+\vk\cdot\vp+\vp^2-\MN E-\ii\eps} \\
 +\int\ddq\,\Delta_d\left(E-\frac{\vq^2}{2\MN},\vq\right)
 \cdot(\ii\TF^{\mathrm{a},ik})^{\gamma c}_{\alpha a}(E;\vk,\vq)
 \cdot\frac{\MN\yd\yt}{2}\frac{(\sigma^k)^\beta_\gamma(\tau^B)^b_c}
 {\vq^2+\vq\cdot\vp+\vp^2-\MN E-\ii\eps} \\
 +\int\ddq\,\Delta_t\left(E-\frac{\vq^2}{2\MN},\vq\right)
 \cdot(\ii\TF^{\mathrm{b},iC})^{\gamma c}_{\alpha a}(E;\vk,\vq)
 \cdot\Biggl\{\frac{\MN\yt^2}{2}\frac{\delta^\beta_\gamma(\tau^B\tau^C)^b_c}
 {\vq^2+\vq\cdot\vp+\vp^2-\MN E-\ii\eps} \\
 +\alpha\MN^2\yt^2\cdot\delta^\beta_\gamma
 \left(\delta^{CB}-\ii\leviciv^{3CB}\right)
 \left(\frac{\one+\tau_3}{2}\right)^b_c
 \left[\frac{\mathcal{I}_\mathrm{bubble}(E;\vq,\vp)}{(\vq-\vp)^{2}+\lambda^2}
 -\frac{\rd}{2}\frac{1}{(\vq-\vp)^2+\lambda^2}\right]\Biggr\}
\label{eq:pd-IntEq-D-b-raw}
\end{multline}
\label{eq:pd-IntEq-D-raw}
\end{subequations}
for the $p$--$d$ equation depicted in Fig.~\ref{fig:pd-IntEq-D}.

\section{Normalisation of the trinucleon wave functions}
\label{sec:Triton-Norm}

In this section we will give a brief derivation of the normalisation
condition~\eqref{eq:Triton-WF-norm} for the triton wave functions used in
Sec.~\ref{sec:He-3}. In order to do that we need to introduce a little more
formalism. For simplicity, we work with a simplified nucleon--deuteron system,
where the virtual spin-singlet state is neglected. The discussion could easily
be carried out for the full coupled-channel system, but that would only
complicate the notation.

\subsection*{Bethe--Salpeter equation}

We start by considering the full two-body nucleon--deuteron propagator (Green's
function) $G$, which fulfils the (inhomogeneous) Bethe--Salpeter
equation\footnote{For a discussion of the Bethe--Salpeter we refer
to~\cite{Lurie}, which served as a starting point for our considerations.} in
momentum space:
\begin{equation}
 G(k,p;P) = G_0(k,p;P) + \int\dfq{q}\int\dfq{q'}\,G(k,q;P)
 \cdot K(q,q';P)\cdot G_0(q',p;P) \,.
\label{eq:BS-ms}
\end{equation}
$G_0$ is essentially a product of a nucleon propagator $\Delta_N$ and a deuteron
propagator $\Delta_d$. More precisely, we have
\begin{equation}
 G_0(k,p;P) = (2\pi)^4\fvdelta(k-p)\cdot\Delta_d\left(\eta_dP+p\right)
 \cdot\Delta_N\left(\eta_NP-p\right) \,,
\label{eq:G0-ms}
\end{equation}
where $P$ is the total four-momentum of the system and $\eta_d+\eta_N=1$. $K$
represents the (doublet-projected) one-nucleon exchange diagram,
\begin{equation}
 K(k_0,\vk,p_0,\vp;E) = \frac{-{\ii\yd^2/2}}
 {\eta_dE - \eta_NE + k_0 + p_0-\frac{(\vk+\vp)^2}{2\MN}+\ii\eps} \,,
\label{eq:1N-exchange}
\end{equation}
as shown, for example, in Fig.~\ref{fig:nd-IntEq-Q}.

Assuming the existence of a trinucleon bound state (the triton in our current
toy model) at an energy $E=-{E_B}<0$, one can show that
\begin{equation}
 G(k,p;P)=\ii\frac{\psi_{B\vP}(p)\psi^\dagger_{B\vP}(k)}{E+E_B+\ii\eps}
 \ +\ \text{terms regular at $P_0 = E = -{E_B}$} \,,
\label{eq:G-factorisation}
\end{equation}
\ie\ $G$ factorises at the bound state pole.

\subsection*{Three-dimensional reduction}

We now consider a bound state at rest, \ie~$P=(-{E_B},\vZero)$, and define the
\emph{amputated} wave function
\begin{equation}
 \Bgen(p_0,\vp)=\psi_{B\vZero}(p_0,\vp)
 \cdot\left[\Delta_d\left(-{\eta_dE_B}+p_0,\vp\right)\right]^{-1}
 \cdot\left[\Delta_N\left(-{\eta_NE_B}-p_0,\vp\right)\right]^{-1} \,,
\label{eq:BS-amputated}
\end{equation}
which fulfils the homogeneous equation
\begin{equation}
 \Bgen(p_0,\vp)=\int\dfq{q}\,K(q,p;-{E_B})
 \cdot\Delta_d\left(-{\eta_dE_B}+q_0,\vq\right)
 \cdot\Delta_N\left(-{\eta_NE_B} -q_0,\vq\right)
 \cdot\Bgen(q_0,\vq) \,.
\label{eq:BSE-hom-MS-3}
\end{equation}
Carrying out the $\dd q_0$-integration picks up the residue from the nucleon
propagator pole at $q_0 = -{\eta_NE_B}-\vq^2/(2\MN)+\ii\eps$. From the resulting
right hand side of~\eqref{eq:BSE-hom-MS-3} we then find that
\begin{equation}
 \Bgen(\vp)\equiv\Bgen\left(-{\eta_NE_B}-\frac{\vp^2}{2\MN},\vp\right)
\label{eq:BS-amputated-2}
\end{equation}
fulfils the equation
\begin{equation}
 \Bgen(\vp)=\int\ddq\,
 K\left(\eta_NE-\frac{\vq^2}{2\MN},\vk,\eta_NE-\frac{\vp^2}{2\MN},\vp;E\right)
 \cdot\Delta_d\left(-{E_B}-\frac{\vq^2}{2\MN},\vq\right)\cdot\Bgen(\vq) \,.
\label{eq:BSE-hom-MS-amp}
\end{equation}
This is essentially the single-channel equivalent of~\eqref{eq:BS-IntEq} (before
S-wave projection), so we have established the connection of our current
formalism to the triton wave functions $\vec{\BS}$ in Sec.~\ref{sec:He-3}. Note
furthermore that
\begin{equation}
 \phi(\vp) \equiv \int\dn{p}\;\psi(p_0,\vp)
 = \Delta_d\left(-{E_B}-\frac{\vp^2}{2\MN},\vp\right)\cdot\Bgen(\vp)
\label{eq:phi}
\end{equation}
is a Schrödinger wave function. We now write~\eqref{eq:BS-ms} in an operator
notation as
\begin{equation}
 G = G_0+GKG_0 = G_0+G_0KG \,.
\label{eq:BS-op}
\end{equation}
Defining
\begin{equation}
 \sktilde{G}(\vk,\vp;-{E_B}) = \int\dn{k}\int\dn{p}\;G(k,p;P) \,,
\end{equation}
we find
\begin{equation}
 \sktilde{G}\sim\ii\frac{\ket{\phi}\bra{\phi}}{E+E_B}
 \,\mathtext{for}\,E \to -{E_B} \,,
\end{equation}
where $\ket{\phi}$ corresponds to the wave function given in~\eqref{eq:phi}, and
\begin{equation}
 \sktilde{G} = \sktilde{G}_0+\sktilde{G_0KG} \,.
\end{equation}
From this we readily derive the normalisation condition
\begin{equation}
 \ii\bra{\phi}{\frac{\dd}{\dd E}\left({\sktilde{G}_0}^{-1}
 -\sktilde{V}\right)}\ket{\phi}\Big|_{E = -{E_B}}=1 \,,
\label{eq:Norm-3dBS}
\end{equation}
where
\begin{equation}
 \sktilde{V} \equiv {\sktilde{G}_0}^{-1} - {\sktilde{G}}^{-1} \,.
\label{eq:V-3dBS}
\end{equation}
A straightforward calculation shows that
\begin{equation}
 {\sktilde{G}_0}^{-1}(\vk,\vp;E) = (2\pi)^3\vdelta(\vk-\vp)
 \cdot\left[\Delta_d\left(E-\frac{\vp^2}{2\MN},\vp\right)\right]^{-1} \,,
\label{eq:G0m1-3dBS}
\end{equation}
and we also see that $\ket{\Bgen} = {\sktilde{G}_0}^{-1}\ket{\phi}$. The
expression for $\sktilde{V}$ \textit{a priori} looks more complicated, but one
finds that in the formal expansion
\begin{equation}
 \sktilde{V} = {\sktilde{G}_0}^{-1}
 -\left[\sum\limits_{n=0}^\infty\left(-{\sktilde{G}_0}^{-1}\,
 \sktilde{G_0KG}\right)^n\right]{\sktilde{G}_0}^{-1}
\label{eq:V-3dBS-exp}
\end{equation}
everything but the term
\begin{equation}
 \sktilde{V}_1 \equiv {\sktilde{G}_0}^{-1}\sktilde{G_0KG_0}{\sktilde{G}_0}^{-1}
\end{equation}
drops out, and we have
\begin{equation}
 \sktilde{V}_1(\vk,\vp;E) = K\left(\eta_NE-\frac{\vk^2}{2\MN},\vk,\eta_NE
 -\frac{\vp^2}{2\MN},\vp;E\right) \,.
\end{equation}
The essential ingredient to see this is
\begin{equation}
 \sktilde{\cdots G_0}{\sktilde{G}_0}^{-1}\sktilde{G_0\cdots}
 = \sktilde{\cdots G_0\cdots} \,,
\end{equation}
which, in turn, follows from the fact that the nucleon propagator residues are
always picked up in such a way that one deuteron propagator is cancelled by the
inverse propagator in ${\sktilde{G}_0}^{-1}$, \cf\ Eq.~\eqref{eq:G0m1-3dBS}.
Altogether, we have shown that~\eqref{eq:Norm-3dBS} is just the
single-channel version of the normalisation condition~\eqref{eq:Triton-WF-norm}
stated in Sec.~\ref{sec:He-3} (modulo S-wave projection), where the functions
$I_{d,t}$ correspond to ${\sktilde{G}_0}^{-1}$.

\end{document}